\documentclass[letterpaper]{article}
\usepackage{amssymb,amsmath,amsthm,bbm}

\newcommand{\eq}[1]
    {\begin{equation}
        #1
     \end{equation}}

\newcommand{\spliteq}[1]
    {\begin{equation}\begin{split}
	    #1
	 \end{split}\end{equation}}
	 
\newcommand{\Res}{{\rm Res}}

\newcommand{\e}{{\rm e}}
\renewcommand{\d}{{\rm d}}
\renewcommand{\i}{{\rm i}}
\renewcommand{\Im}{{\rm Im}}
\newcommand{\brac}[1]{\left(#1\right)} 
\newcommand{\set}[1]{\left\{#1\right\}}
\newcommand{\ebrac}[1]{\left[#1\right]} 
\newcommand{\abs}[1]{\left\arrowvert #1\right\arrowvert}
 
\newcommand{\ket}[1]{\left|#1\right>}
\newcommand{\norm}[1]{\left<#1\left|#1\right>\right.} 

\newcommand{\cor}[1]{\left<#1\right>}

 \newcommand{\sgn}{\mathrm{sgn}}

\begin{document}

\begin{titlepage}
\setcounter{page}{0}
\begin{flushright}
      arXiv:0801.2711 [hep-th]\\
      ITP--UH--02/08\\
\end{flushright}
\vspace{20mm}
\begin{center}
{\LARGE\bf On Factorization Constraints for Branes in the $\bf H_3^+$ Model}\\
\vspace{5mm}
{\sc Hendrik Adorf} and {\sc Michael Flohr}\\
\vspace{5mm}
{\sl Institut f\"ur Theoretische Physik,}\\
{\sl Gottfried Wilhelm Leibniz Universit\"at Hannover,}\\
{\sl Appelstra\ss e 2, 30167 Hannover, Germany.}\\
\vspace{5mm}
{\sc e-mail:} \tt adorf, flohr@itp.uni-hannover.de\\
\vspace{5mm}
{\small January 17, 2008}
\end{center}
\vspace{10mm}
\begin{abstract}
\noindent We comment on the brane solutions for the boundary $\rm H_3^+$ model that have been proposed so far and point out that they should be distinguished according to the patterns regular/irregular and discrete/continuous. In the literature, mostly irregular branes have been studied, while results on the regular ones are rare. For all types of branes, there are questions about how a second factorization constraint in the form of a $b^{-2}/2$-shift equation can be derived. Here, we assume analyticity of the boundary two point function, which means that the Cardy-Lewellen constraints remain unweakened. This enables us to derive unambiguously the desired $b^{-2}/2$-shift equations. They serve as important additional consistency conditions. For some regular branes, we also derive $1/2$-shift equations that were not known previously. Case by case, we discuss possible solutions to the enlarged system of constraints. We find that the well--known irregular continuous $AdS_2$ branes are consistent with our new factorization constraint. Furthermore, we establish the existence of a new type of brane: The shift equations in a certain regular discrete case possess a non--trivial solution that we write down explicitly. All other types are found to be inconsistent when using our second constraint. We discuss these results in view of the Hosomichi--Ribault proposal and some of our earlier results on the derivation of $b^{-2}/2$-shift equations.
\end{abstract}
\vfill
\end{titlepage}

%\phantom{A}
%\vspace{10mm}
%\hrule
\setcounter{section}{-1}
\section{Preamble}
This article reconsiders the results that we had published in \cite{Adorf:VariousH3Branes}. In the course of revising \cite{Adorf:VariousH3Branes}, it turned out that some of its formulae and statements had been misleading, due to some subtleties in the analytic continuations that we have to use. Now, after a thorough revision, both our results and our viewpoint on the whole subject have changed and the paper itself has grown and changed immensely. We have therefore decided to publish it as a completely new article (which it actually is, since only some parts of the introductory material and of the appendices have stayed unaltered) rather than merely replacing the old one. {\bf This work therefore supersedes \cite{Adorf:VariousH3Branes}}, as it corrects the misleading formulae and statements and puts everything into a new perspective. Nonetheless, it is our decision to leave \cite{Adorf:VariousH3Branes} in its old form on the arxive, because the basic ideas of exploring certain patterns systematically and trying to treat the $\rm H_3^+$ boundary two point function analytically are already formulated there.
%\vskip .5cm
%\hrule

%\newpage
\section{Introduction}
The $\rm H_3^+$ model, which is a suggestive way to denote the ${\rm SL}(2,\mathbb{C})/{\rm SU}(2)$ WZNW model, has been studied for quite some time now, the motivations being at least fourfold: On the one hand, it falls into the class of non--compact conformal field theories (CFTs) whose general structure and features are very poorly understood so far. On the other hand, it is essential for a study of the bosonic string in certain curved backgrounds. While the $\rm H_3^+$ model itself describes the bosonic string in an euclidean $AdS_3$ background, it can be analytically continued to the lorentzian $AdS_3$ string. The latter is of great interest, particularly in view of the ${\rm AdS}_3/{\rm CFT}_2$ correspondence. See references \cite{GKS2, KS, BORT, SAT1, SAT2, GN, GNA, SR2, Dan1, MaldaOoguri1, MaldaOoguri2, MaldaOoguri3} and further references therein. Thirdly, from the euclidean $AdS_3$ string there is a connection to the so-called cigar CFT \cite{RibScho}, which describes a bosonic string moving in an euclidean 2D black hole \cite{WIT, DVV, BB, HPT}. Finally, a forth reason to study the $\rm H_3^+$ model is its very interesting duality to Liouville theory \cite{JT3, ZAF}, which has been remarkably generalized in \cite{TeschRi} and \cite{HikidaSchomer:H3Liouville} and extended to the CFT with $AdS_2$ boundary in \cite{HoRi,Ribault:AdSBdry3PtFct}.\\
Concerning the bulk $\rm H_3^+$ model, its structure is apparently quite well explored (see \cite{JT3, JT1, JT2} and \cite{BP1}), although some subtleties still persist (e.g. \cite{BP1, GIRI1, GIRI2, NIC1}). Looking at the corresponding boundary CFT, we find that the picture is rather more incomplete. In particular, the question of what branes can consistently be described does not seem to be fully answered up to now. One approach to this issue, that has been pursued in \cite{PST, SR1, GKS, PS} and \cite{LOP} is to compute boundary one point functions. These are actually fixed to great extent by boundary Ward identities. Their only remaining degree of freedom is the so-called one point amplitude. This is an interesting object to study, because it describes the coupling of a closed string in the bulk to a D-brane. Accordingly, it must depend on the properties of these two objects. Seeing that closed strings are characterized by an ${\rm sl}(2,\mathbb{C})$-'spin' label $j$ (see chapter \ref{Review}) and branes are labelled by a complex parameter $\alpha$, a one point amplitude is denoted $A(j\arrowvert\alpha)$.\footnote{It can also depend on some more data, see chapters \ref{BraneTypes} and \ref{irrAdS2d-rho2}.} In the sequel, when talking about a brane solution, we actually mean a solution for the one point amplitude. For other aspects and further references concerning the boundary $\rm H_3^+$ model, we refer the reader to the lecture notes \cite{VS1} and \cite{VS2}.

\subsubsection*{\sl Approach used in the present Article}
The strategy in the computation of one point amplitudes is to derive a consistency condition (a so--called shift equation) for them and then try to solve it. Such constraints have been studied for general rational CFTs by Cardy and Lewellen \cite{CardyLewellen:BulkBdryOps, Lewellen:Sewing}. The nature of a shift equation is to relate the one point amplitude for some string label $j$ to a sum of one point amplitudes taken at shifted string labels like e.g. $j\pm 1/2$. See equation (\ref{irrAdS2d-rho2-Shift1}) for an example. Generically, a solution for the one point amplitude will not exist for arbitrary boundary conditions, but restrictions will apply. By the same token, the labels $j$ of strings that do couple consistently are expected to be constrained.\\
In order to derive a shift equation, a special two point function involving one degenerate field is considered (see chapter \ref{Review} for an explanation of the term 'degenerate field'). The benefit of using a degenerate field here is, that it allows to solve for the two point function exactly. The shift equation is then extracted from it by taking a factorization limit in that the two point correlator factors into two one point correlators. The simplest case, from which a $1/2$-shift equation descends, uses a degenerate field with ${\rm sl}(2,\mathbb{C})$-'spin' label $j=1/2$. The solution to only one such shift equation is however not unique. Unfortunately, for the most important cases, the existing literature only provides this $1/2$-shift equation and proposes a one point amplitude that solves it. In order to fix the solutions uniquely and hence back up their consistency, a second independent shift equation is desirable.\\
The natural candidate from which to derive that second factorization constraint is the boundary two point function involving the next simple degenerate field which has ${\rm sl}(2,\mathbb{C})$ label $j=b^{-2}/2$.
The aim of the present article is to study this boundary two point function, analyze how it provides us with the desired $b^{-2}/2$-shift equation and study the implications of the new constraints. To this end, we define the correlator as a solution to the Knizhnik-Zamolodchikov equation that it has to obey. This solution is not everywhere defined, but only in a certain region of the real $(u,z)$-plane\footnote{The conformal fields of the $\rm H_3^+$ theory depend on two complex variables: A space-time coordinate $z$ and an internal variable $u$ - see chapter \ref{Review}. The real $(u,z)$-plane we talk about here, is the plane spanned by the real-valued crossing ratios formed from internal ($u$) and space-time ($z$) positions of the fields in the boundary CFT correlator.}. Unfortunatley, in order to take the factorization limit, one has to move out of this initial region. We therefore need to extend the definition of the boundary two point function under consideration to other domains in the $(u,z)$-plane. Since, in its initial domain, it is in fact an analytic function of both variables $(u,z)$, we shall assume that it can be extended to other regions by analytic continuation in $(u,z)$. 

\subsubsection*{\sl The Hosomichi--Ribault Proposal}
General $\rm H_3^+$ boundary correlators have been studied recently in \cite{HoRi} by using a mapping to Liouville theory. In formulating the mapping, it was necessary to distinguish between two non-overlapping regimes, namely the bulk and the boundary regime. Thus, the question arose how correlators should behave when moving from one regime into the other. The proposal of \cite{HoRi} is that correlators should have a finite limit and be continuous at the interface of the two regimes. With this modest requirement, the authors expect a weakening of the Cardy-Lewellen factorization constraint.\\
Now, our assumption of analyticity seems to circumvent the Hosomichi-Ribault proposal. Essentially, the region where our boundary two point function is defined initially lies in the bulk regime and the region where the factorization limit is taken lies in the boundary regime. So indeed, assuming an analytic boundary two point function, no interface at which the correlator behaves distinguishedly different from anywhere else would be singled out. Moreover, and even more importantly, the Cardy--Lewellen constraints remain unweakened.\\
Nevertheless, when attempting to understand the $\rm H_3^+$ branes in a model-intrinsic way, i.e. without using any mapping to a different theory, there is a priori no reason why the boundary two point correlator might behave 'unusual' anywhere. Especially when recalling that our boundary two point function under consideration is explicitely given as an analytic function in its initial domain, it seems very natural that it can be recovered in other domains by analytic continuation.\\
Hence, the point of view taken up in this paper is to forget for a moment about any information that comes from outside the $\rm H_3^+$ model and see how far our treatment can carry us. If the Hosomichi-Ribault proposal is correct, our viewpoint has to break down at some stage. We would like to learn where and how this might happen.\\
Yet, the features that we find are rather nice: Our assumption enables us to derive the desired $b^{-2}/2$-shift equations for boundary $\rm H_3^+$ one point functions. Marvellously, the well-known irregular continuous $AdS_2$ branes remain consistent even with this new constraint. Moreover, we even discover a different and, to our knowledge, new kind of brane. We also take a closer look at the analytically continued boundary two point function and discuss in detail how the analytic continuations are taken, carefully taking domains of convergence and branch cuts of the occuring hypergeometric and generalized hypergeometric functions into account. We argue that the two point function shows the features expected from the Hosomichi--Ribault proposal (except for the unweakened Cardy--Lewellen constraint), which are finiteness at $u=z$ and continuity at $u=z$ in the real $(u,z)$-plane. Additionally, as expected from \cite{GIRI1}, our two point function shows logarithms in the patch $z>u$. We argue that these logarithms are merely "coordinate singularities" that cannot be interpreted as coming from logarithmic OPEs: The bulk-boundary OPE stays free of logarithms, so that the $\rm H_3^+$ CFT does not appear to constitue a logarithmic CFT \cite{Flohr:BitsAndPieces}.\\
What we would like to stress here is, that our assumption of analyticity of the boundary two point function implies some caveats when comparing our derivations and results with the literature that focuses on the continuity proposal. First of all, the derivations are done in the space $\mathbb{C}^2$, i.e. for complex valued $u,z$. Thus, while working analytically, we have to free ourselves from any connotations that suggest a connection between the occuring $u$ and $\bar{u}$. During analytic continuation, $\bar{u}$ is to be read as just another chiral (!) variable and the relation between $u$ and $\bar{u}$ is just the same as the relation between two chiral variables $u_1$ and $u_2$, say. Therefore, expressions such as $(u+\bar{u})^{2j}$ are generally not of the form $|u+\bar{u}|^{2j}f(\sgn(u+\bar{u}))$, even not for $2j$ an integer. Thus, we have to consider them individually. Only after the analytic continuation is done do we take the real cut where $\bar{u}$ is identified with the complex conjugate of $u$, as this is the region of physical significance in our context. There are many more pitfalls, and we urge the reader to be very careful when carrying over common assumptions or facts from the body of literature making use of the continuity proposal, which applies for $\mathbb{R}^2$, i.e. real valued $u,z$, only.

\subsubsection*{\sl Regular\footnotemark and Irregular\addtocounter{footnote}{-1}\footnote{We are going to introduce this terminology in section \ref{Regular-Irregular}.} Branes}
Let us now account briefly for the different kinds of brane solutions that are found in the existing literature. In \cite{PST}, the authors showed that there are two classes of branes: $AdS_2$ and $S^2$ branes. They derived one shift equation for each class and also proposed solutions. Afterwards, \cite{SR1} enlarged the picture and introduced the so-called $AdS^{(d)}_2$ branes, $(d)$ standing for {\sl discrete}. The author of \cite{SR1} was guided by some relation between the ZZ and FZZT branes of Liouville theory that, in the spirit of the Liouville/$\rm H_3^+$ correspondence of \cite{TeschRi}, was carried over to the $AdS_2$ branes of \cite{PST}. However, we like to point out that these new branes can also be understood as arising from the following difference in the derivation of the shift equation: The degenerate field is always expanded in terms of boundary fields, using its bulk--boundary OPE. Now, assuming a discrete open string spectrum on the brane, the occuring bulk--boundary OPE coefficient that corresponds to propagation of the identity in the open string channel, can be identified with the one point amplitude. Hence, the two point function factorizes into a product of two one point functions. On the other hand, assuming a continuous open string spectrum, the above identification is lost. Instead, the two point function becomes a product of a one point function and a residue of the bulk--boundary OPE coefficient corresponding to the identity propagation. This is explained in \cite{VS2} and we review it in section \ref{Discrete-Continuous}. The first case results in the $AdS_2^{(d)}$, whereas the second case leads to the $AdS_2^{(c)}$ shift equations, $(c)$ standing for continuous. This treatment can always be applied, no matter what gluing condition we are using. This has actually been recognized, but not fully exploited, by the authors of \cite{GKS}.\\
Besides this scheme, that we think should be employed more systematically, there seems to be another pattern that has not been taken much care of up to now. In \cite{GKS}, a solution to the boundary conformal Ward identities for the one point function, that is everywhere regular in the internal variable $u$ (see chapter \ref{Review}), was proposed. Opposed to this solution, \cite{PST, SR1} and \cite{LOP} use a one point function that is not everywhere regular. While both solutions are correct (see chapter \ref{BraneTypes}), we find that they give rise to slightly different shift equations (see chapters \ref{irrAdS2d-rho1}, \ref{regAdS2d-rho2} and \ref{regAdS2d-rho1} in case of the discrete and \ref{irrAdS2c-rho2-1}, \ref{regAdS2c-rho2} and \ref{regAdS2c-rho1} for the continuous branes). The modifications that arise for the regular dependence opposed to the irregular one, change the qualitative behaviour of possible solutions significantly. Consequently, not only should one distinguish between continuous and discrete, but also between regular and irregular D-brane solutions. In section \ref{regAdS2d-rho1} we demonstrate that a consistent non--trivial solution for certain regular discrete branes exists.

\subsubsection*{\sl Plan of the Paper}
The paper is organized as follows: After having fixed some notation in chapter \ref{Review}, we elaborate on the distinction between continuous and discrete branes as well as regular and irregular one point functions in chapters \ref{GlueConds} and \ref{BraneTypes}. This is followed by the derivation of the shift equation involving degenerate field $\Theta_{b^{-2}/2}$ for an irregular $AdS_2^{(d)}$ brane in chapter \ref{irrAdS2d-rho2}. In that process an analytic continuation of the boundary two point function is needed. We show that the $AdS_2^{(d)}$ solution that had been proposed earlier does not solve our new shift equation. Then we go on to derive and discuss the solutions to the shift equations (involving degenerate fields with ${\rm sl}(2,\mathbb{C})$-'spin' labels $j=1/2$ and $j=b^{-2}/2$ respectively) for the remaining discrete branes in chapters \ref{irrAdS2d-rho1}, \ref{regAdS2d-rho2} and \ref{regAdS2d-rho1}. The only consistent case is the one of regular discrete branes. Afterwards, in chapters \ref{irrAdS2c-rho2-1}, \ref{regAdS2c-rho2} and \ref{regAdS2c-rho1}, we give $1/2$- and $b^{-2}/2$-shift equations for the various continuous branes and also comment on their possible solutions. Here, our shift equation confirms the consistency of the irregular $AdS_2^{(c)}$ branes of \cite{PST}. Afterwards, in chapter \ref{CloseLook}, we take a closer look at our analytically continued two point function and discuss some of its features, including details of the analytic continuation process. Finally, we summarize our results in chapter \ref{Conclusion}, where we also suggest further directions and discuss open questions. The more technical calculational details and some useful formulae can be found in various appendices.

\section{\label{Review}A Brief Review of the Bulk $\bf H_3^+$ Model}
The bulk $\rm H_3^+$ model has been fairly well studied, see \cite{JT3, JT1, JT2} and \cite{BP1}. Here, we essentially fix our notation (which follows very closely \cite{PST}) and summarize those facts and formulae which will be indispensable in the sequel. They can all be found in \cite{JT1, JT2} and \cite{PST}.\\
Besides conformal symmetry, the $\rm H_3^+$ model possesses an affine $\hat{\rm sl}(2,\mathbb{C})_k\times\hat{\rm sl}(2,\mathbb{C})_k$ symmetry, i.e. its chiral algebra does not only consist of an energy momentum tensor $T(z)$, but also of the currents $J^a(z)=\sum_n z^{-n-1}J^a_n$, $a\in\set{+,-,3}$ (plus a corresponding antichiral sector). Primary fields fall into representations of the zero mode algebra (generated by the operators $J^a_0$) and are henceforth labelled by a pair of ${\rm sl}(2,\mathbb{C})$-'spins' $(j,\bar{j})$, and a pair of internal variables, which will be denoted by $(u,\bar{u})\in\mathbb{C}^2$, so that a typical primary field should be denoted $\Theta_{j,\bar{j}}(u,\bar{u}|z,\bar{z})$. However, from now on we will always suppress the barred variables. The $\hat{\rm sl}(2,\mathbb{C})_k$-currents act on these primaries via the operator product expansion (OPE)
\eq{J^a(z)\Theta_j(u|w)=\frac{D^a_j(u)\phi_j(u|w)}{z-w},}
i.e. the zero mode algebra is represented through the differential operators $D^a_j(u)$, given by
\eq{
D^+_j(u)=-u^2\partial_u+2ju, \hskip .5cm 
D^-_j(u)=\partial_u, \hskip .5cm 
D^3_j(u)=u\partial_u-j.}
Analogous formulae hold for the antichiral sector. Through the standard Sugawara construction, the energy momentum tensor is expressed in terms of products of the currents and thereby a relation between conformal weight $h$ and 'spin'-label $j$ of primary fields is established:
\eq{h\equiv h(j)=-\frac{j(j+1)}{k-2}=-b^2 j(j+1)\,,}
which implicitly defines the relation between the parameter $b$ and the affine Kac--Moody level $k$. Note that there is a reflection symmetry, namely $h(-j-1)=h(j)$. This leads one to identify the representations with labels $j$ and $-j-1$ and gives rize to a relation between primary fields $\Theta_j(u|z)$ and $\Theta_{-j-1}(u|z)$:
\eq{\label{RefSymm}\Theta_j(u|z)=-R(-j-1)\frac{2j+1}{\pi}\int_{\mathbb{C}}{\rm d}^2 u'|u-u'|^{4j}\Theta_{-j-1}(u'|z),}
where the {\sl reflection amplitude} $R(j)$ is given by
\eq{\label{Rj}R(j)=-\nu_b^{2j+1}\frac{\Gamma(1+b^2(2j+1))}{\Gamma(1-b^2(2j+1))}.}
The physical spectrum (normalisable operators) of the bulk theory consists of the so-called continuous representations \cite{JT2}, that are parametrized through $j\in -\frac{1}{2}+\i\mathbb{R}_{\geq 0}$ and are in fact infinite dimensional.\\
By the usual operator-state correspondence, to each primary field $\Theta_j$ corresponds a highest weight state $\ket{j}$. It has the property that $J^a_n\ket{j}=0$ for all $n>0$. Acting on it with the $J^a_{n<0}$ generates a whole Verma module $V_j$. These modules are reducible, iff
\eq{j=j_{r,s}:=-\frac{1}{2}+\frac{1}{2}r+\frac{b^{-2}}{2}s,}
where either $r\geq 1$, $s\geq 0$ or $r<-1$, $s<0$ (see \cite{JT1}). This means that they possess null-submodules. These are submodules that are generated by so-called {\sl null states} (or {\sl singular vectors}), i.e. states $\ket{\mathrm{null}}$ with $\norm{\mathrm{null}}=0$. Those primary fields $\Theta_{j_{r,s}}$ that give rise to reducible modules are called {\sl degenerate fields}. In order to get an irreducible module out of a reducible one, all null-submodules have to be divided out of the original module. This in turn gives rise to certain differential equations, that all correlators involving the corresponding degenerate field have to solve. In this paper, we shall make use of the degenerate fields $\Theta_{j_{r,s}}$ associated to $j_{2,0}=1/2$ and $j_{1,1}=b^{-2}/2$.

\section{Boundary $\bf H_3^+$}
In this section, we have several comments to make on the boundary CFT that one obtains from the $\rm H_3^+$ model. Specifically, we discuss the various kinds of branes that we think should a priori be carefully distinguished.

\subsection{\label{GlueConds}Gluing Conditions}
We choose maximal symmetry preserving boundary conditions. This is done by imposing a gluing condition along the boundary (which is taken to be the real axis)
\eq{J^a(z)=\rho^{a}_{\hphantom{a}b}\bar{J}^b(\bar{z}) \hskip .5cm {\rm at} \hskip .1cm z=\bar{z},}
where $\rho$ is the 'gluing map' i.e. an automorphism of the chiral algebra which leaves the Virasoro field invariant. Thus, by the Sugawara construction, we also have
\eq{T(z)=\bar{T}(\bar{z}) \hskip .5cm {\rm at} \hskip .1cm z=\bar{z},}
and hence not only is a subgroup of the current algebra symmetry preserved, but also half of the conformal symmetry. In the case of ${\rm SL}(2)$ there are four possible gluing maps $\rho_1,\dots,\rho_4$: 
\spliteq{\label{Gluing}
\rho_1\bar{J}^3=\bar{J}^3 \hskip .5cm 
&\rho_1\bar{J}^{\pm}=\bar{J}^{\pm},\\
\rho_2\bar{J}^3=\bar{J}^3 \hskip .5cm 
&\rho_2\bar{J}^{\pm}=-\bar{J}^{\pm},\\
\rho_3\bar{J}^3=-\bar{J}^3 \hskip .5cm
&\rho_3\bar{J}^{\pm}=\bar{J}^{\mp},\\
\rho_4\bar{J}^3=-\bar{J}^3 \hskip .5cm
&\rho_4\bar{J}^{\pm}=-\bar{J}^{\mp}.\\
}                                                         
For now, we will only be concerned with the first and second case, $\rho_1$ and $\rho_2$. The branes associated to $\rho_2$ are conventionally called $AdS_2$ branes \cite{PST}.

\subsection{\label{BraneTypes}Various Types of Branes}
In our study of the $\rm H_3^+$ branes, we will distinguish between discrete and continuous as well as regular and irregular branes. The adjectives discrete and continuous allude to the open string spectra an the branes, whereas regular and irregular refer to the $u$-dependence of the one point functions. We elaborate on these notions in the next two subsections.

\subsubsection{\label{Discrete-Continuous}Discrete and Continuous Branes}
For each of the above four classes of boundary conditions, one can obtain at least two different brane solutions: The 'continuous' and the 'discrete' branes. By the term 'brane solution' we mean the one point amplitude of a generic field $\Theta_j$ in the presence of some boundary condition. The characterising adjectives 'continuous' and 'discrete' relate to the parameter spaces of these solutions or, equivalently, to the open string spectra on the branes. For example, in \cite{PST}, a solution for the {\sl continuous} $AdS_2$ branes was proposed, whereas \cite{SR1} proposed a solution for the {\sl discrete} $AdS_2$ branes. From now on, we will carefully distinguish these different kinds of solutions, by adding a superscript $(c)$ in case of a continuous brane and $(d)$ for a discrete one, as it has already been done in \cite{SR1}. Let us now explain where the difference between continuous and discrete branes originates and how it leads to different factorization constraints. For convenience, let us fix the gluing map to be $\rho=\rho_2$. The discussion for $\rho_1$ is completely analogous.\\
Assuming a discrete open string spectrum on the brane, the bulk-boundary OPE for $\Theta_{j_{r,s}}$ is
\spliteq{\Theta_{j_{r,s}}(u_2|z_2)&=\sum_{\set{l_0}}\abs{z_2-\bar{z}_2}^{-2h(j_{r,s})+h(l_0)}
\abs{u_2+\bar{u}_2}^{2j_{r,s}+l_0+1}\cdot\\
&\cdot C_{\sigma}(j_{r,s},l_0|\alpha)\brac{{\cal J}\Psi}^{\alpha\,\alpha}_{l_0}\brac{u_2\left|{\rm Re}(z_2)\right.}\set{1+{\cal O}(z_2-\bar{z}_2)}\,,}
where $\set{l_0}$ is a discrete set of ${\rm SL}(2)$-'spin' labels, $\Psi^{\alpha\,\alpha}_{l}(t|x)$ is a primary boundary field ($t,x\in\mathbb{R}$) and we have defined 
\eq{({\cal J}\Psi)^{\alpha\,\alpha}_{l}(u|x):=\int_{\mathbb{R}}\frac{\d t}{2\pi} \abs{u+\i t}^{-2l-2}\Psi^{\alpha\,\alpha}_{l}(t|x)\,.}
Note that under a scaling $u\mapsto \lambda u$, this transforms as
\eq{({\cal J}\Psi)^{\alpha\,\alpha}_{l_0}(\lambda u|x)=\lambda^{-l_0-1}({\cal J}\Psi)^{\alpha\,\alpha}_{l_0}(u|x)\,,}
so that the scaling properties of $\Theta_{j_{r,s}}(u_2|z_2)$ on the L.H.S are matched correctly. Now, the kind of factorization constraint we are seeking for arises when looking at the identity contribution of the bulk-boundary OPE. The corresponding bulk-boundary OPE coefficient $C_{\sigma}(j_{r,s},0|\alpha)$ can be identified with a one-point amplitude:
\eq{C_{\sigma}(j_{r,s},0|\alpha)=A_{\sigma}(j_{r,s}|\alpha)\,.}
Therefore, starting with a two point function and taking the factorization limit leads, in the discrete case, to a product $A_{\sigma_2}(j_{r,s}|\alpha)A_{\sigma_1}(j|\alpha)$.\\
On the other hand, assuming a continuous open string spectrum on the brane, the bulk-bundary OPE of $\Theta_{b^{-2}/2}$ contains 
\eq{\tilde{c}_{\sigma}(j_{r,s},l_0|\alpha):=\Res_{l=l_0}C_{\sigma}(j_{r,s},l|\alpha)} 
rather than $C(j_{r,s},l_0|\alpha)$. The reason for this is given in \cite{VS2}. Let us summarize it here briefly: Since we are using Teschner's Trick, i.e. we are analytically continuing the field label $j_2$ to the label of a degenerate representation $j_2=j_{r,s}$, we should look at the generic bulk-boundary OPE
\spliteq{\Theta_{j_2}(u_2|z_2)&=\int_{{\cal C}^+}\d l\abs{z_2-\bar{z}_2}^{-2h(j_2)+h(l)}
\abs{u_2+\bar{u}_2}^{2j_2+l+1}\cdot\\
&\cdot C_{\sigma}(j_2,l|\alpha)\brac{{\cal J}\Psi}^{\alpha\,\alpha}_{l}\brac{u_2\left|{\rm Re}(z_2)\right.}\set{1+{\cal O}(z_2-\bar{z}_2)}\,,}
where the contour of integration is ${\cal C}^+:=-\frac{1}{2}+\i\mathbb{R}_{\geq 0}$. Since $j_2=j_{r,s}$ is a degenerate representation, only a discrete set of open string modes is excited in the bulk-boundary OPE of its corresponding field operator. Accordingly, when deforming the contour in the process of analytic continuation, only finitely many contributions $\set{l_0}$ are picked up. They come from poles in the $C_{\sigma}(j_{r,s},l|\alpha)$ that cross the contour of integration. Therefore, not the bulk-boundary coefficients themselves, but only their residua occur. Henceforth, we obtain
\spliteq{\Theta_{j_{r,s}}(u_2|z_2)&=\sum_{\set{l_0}}\abs{z_2-\bar{z}_2}^{-2h(j_{r,s})+h(l_0)}
\abs{u_2+\bar{u}_2}^{2j_{r,s}+l_0+1}\cdot\\
&\cdot\tilde{c}_{\sigma}(j_{r,s},l_0|\alpha)\brac{{\cal J}\Psi}^{\alpha\,\alpha}_{l_0}\brac{u_2\left|{\rm Re}(z_2)\right.}\set{1+{\cal O}(z_2-\bar{z}_2)}\,.}
In the factorization limit, we are looking at the identity contribution again, but this time, the residuum of the appropriate bulk-boundary coefficient does not have an obvious relation to a one-point-amplitude. Thus, in the continuous case, we are left with a product $\tilde{c}_{\sigma_2}(j_{r,s},0|\alpha)A_{\sigma_1}(j|\alpha)$.

\subsubsection{\label{Regular-Irregular}Regular and Irregular Branes}
Moreover, we want to argue that there are even more possible brane solutions, that are distinguished by their regularity behaviour when approaching the boundary in internal $u$-space. Let us explain in detail why this is the case for the example that the gluing map is again $\rho=\rho_2$ (the other cases can clearly be treated in just the same way). It is the Ward identites that fix the $u$-dependence of the one point function $G^{(1)}_{j,\alpha}(u|z):=\cor{\Theta_j(u|z)}_{\alpha}$ in the presence of boundary condition $\alpha$ entirely. The equation for $J^-$ tells us that it is a function of $u+\bar{u}$ only. The equations for $J^3$ and $J^+$ show a singularity at $0=u+\bar{u}=:2u_1$. We therefore have to distinguish two cases. The solution for $u_1>0$ is 
\eq{G^{(1)}_{j,\alpha}(u;u_1>0|z)=(u+\bar{u})^{2j}A^{+}_{j,\alpha}(z)} 
and the one for $u_1<0$ reads 
\eq{G^{(1)}_{j,\alpha}(u;u_1<0|z)=(u+\bar{u})^{2j}A^{-}_{j,\alpha}(z).} 
But notice that we could have equally well written 
\eq{G^{(1)}_{j,\alpha}(u;u_1<0|z)=|u+\bar{u}|^{2j}\tilde{A}^{-}_{j,\alpha}(z),} 
where we have just redefined the "constant": $\tilde{A}^{-}_{j,\alpha}(z)=(-)^{2j}A^{-}_{j,\alpha}(z)$. This seems like a harmless thing to do, but we need to be aware that the $u$ dependence has changed from being regular at $u_1=0$ to irregular. Also, for $j\in -\frac{1}{2}+\i\mathbb{R}$, one must give a definition of $(-)^{2j}$. In this and the next chapter, we will compute the one point amplitudes resulting from both these ans\"atze and find that they are indeed very different in nature. The corresponding branes will be called {\sl regular} or {\sl irregular}, respectively. Whether this is an appropriate and useful nomination remains to be seen. At this point, it is important to note that the regular solution is only applicable if $j\in\frac{1}{2}\mathbb{Z}$, in order to avoid a multivalued amplitude\footnote{One might remark here that these branes decouple from the physical spectrum of closed $\rm H_3^+$ strings, which is $j\in\frac{1}{2}+i\mathbb{R}_{\geq 0}$. We shall comment on this objection in the conclusion.}. Let us also mention that in the literature, both kinds of solutions, regular and irregular ones, have been studied. For example, \cite{PST} and \cite{LOP} look at irregular $AdS^{(c)}_2$ and \cite{SR1} treats irregular $AdS^{(d)}_2$ branes, whereas \cite{GKS} studies regular solutions. But up to now, at least to our knowledge, nobody has pointed out that for every case of boundary condition $\rho_1,\dots\rho_4$, we should actually look for both kinds of solutions. Table \ref{T1} shows how little of the 'landscape' has actually been explored so far. It also shows that, except for one case in \cite{GKS}, it has always been only one consistency condition on which the proposed solutions were based, namely the shift equation for the degenerate field $\Theta_{1/2}$. The solutions to this equation are not unique and at least a second consistency condition should be derived that can fix the solution uniquely. The shift equation for the degenerate field $\Theta_{b^{-2}/2}$ can do this job.
\begin{table}
\begin{tabular}{|c||c|c|c|c|c|} \hline
             &$u$-dependence   
                       &\multicolumn{2}{c|}{shift equation (continuous)}   
                                 &\multicolumn{2}{c|}{shift equation (discrete)}\\
             &         
                       &for $\Theta_{1/2}?$    &for $\Theta_{b^{-2}/2}?$ 
                                 &for $\Theta_{1/2}?$     &for $\Theta_{b^{-2}/2}?$\\ \hline\hline
$\rho_1$     &$|u-\bar{u}|^{2j}$ 
                       &\cite{LOP}             &---
                                 &---                    &---\\ \cline{2-6}
             &$(u-\bar{u})^{2j}$ 
                       &\cite{GKS}             &---
                                 &\cite{GKS}             &\cite{GKS}\\ \hline
$\rho_2$     &$|u+\bar{u}|^{2j}$ 
                       &\cite{PST}             &\cite{Adorf:Continuous}                 
                                 &\cite{SR1}             &\cite{Adorf:Continuous}\\ \cline{2-6}
             &$(u+\bar{u})^{2j}$ 
                       &---                    &---
                                 &---                    &---\\ \hline
$\rho_3$     &$|1-u\bar{u}|^{2j}$ 
                       &---                    &---                   
                                 &---                    &---\\ \cline{2-6}  
             &$(1-u\bar{u})^{2j}$       
                       &---                    &---
                                 &\cite{GKS}             &\cite{GKS}\\ \hline
$\rho_4$     &$(1+u\bar{u})^{2j}$       
                       &---                    &---
                                 &\cite{PST}             &---\\ \hline                      
\end{tabular}
\caption{\label{T1} Classes of D-brane solutions and status of their exploration. \cite{GKS} did not distinguish between amplitudes $A^-$ and $A^+$, which is however inevitable (see text). We are therefore reconsidering their results. In \cite{Adorf:Continuous}, we investigated the derivation of the $b^{-2}/2$ factorization constraint using a continuation prescription for the boundary two point function that differs from the one used here; see also our discussion in the concluding section. Note that only one version of $u$-dependence appears for $\rho_4$, as the expression is always strictly positive.}
\end{table}

\subsubsection{\boldmath$AdS_2$ and \boldmath$S^2$ Branes}
From the $u$ dependencies of the one point functions, we can determine what subgroup of the ${\rm SL}(2,\mathbb{C})$ isospin symmetry is preserved by the varying gluing conditions. Since a primary field $\Theta_j(u|z)$ transforms under an ${\rm SL}(2,\mathbb{C})$ isospin transformation $u\mapsto u':=\frac{au+b}{cu+d}$ as
\eq{\Theta_j(u|z)\mapsto\Theta'_j(u'|z)=|cu+d|^{-4j}\Theta_j(u|z)\,,}
one needs to check for every $u$ dependence which ${\rm SL}(2,\mathbb{C})$ subgroup it preserves up to a factor of $|cu+d|^{-4j}$. The result is that the dependencies $|u\pm\bar{u}|^{2j}$ and $(u\pm\bar{u})^{2j}$ preserve an ${\rm SL}(2,\mathbb{R})$ subgroup and are therefore $AdS_2$ branes, whereas $|1-u\bar{u}|^{2j}$ and $(1\pm u\bar{u})^{2j}$ preserve an ${\rm SU}(2)$ subgroup and are thus $S^2$ branes. The cases of gluing maps $\rho_1$ and $\rho_2$ should therefore be isomorphic, as should be those of $\rho_3$ and $\rho_4$. However, we like to advocate here that such a conclusion, which would suggest to leave half of the gluing maps unstudied, might be drawn too quickly here. Indeed, at least one issue remains unclear here: How can $\rho_3$ and $\rho_4$ belong to isomorphic branes if $\rho_3$ allows the inclusion of a signum $\sigma$ and $\rho_4$ does not (see table \ref{T1})? The answer must be that only further consistency checks (like e.g. the shift equations) will forbid the inclusion of a signum for $\rho_3$\footnote{This is, of course, a speculation. Unfortunately, we do not study the spherical branes in this article.}. We take this as a hint that consistency checks will add to the analysis described in this subsection. We therefore consider both gluing maps, $\rho_1$ as well as $\rho_2$, separately. Interestingly, we shall find that in case of irregular branes the shift equations for both gluing maps are isomorphic, whereas in the regular case there are crucial differences (see sections \ref{regAdS2d-rho2} and \ref{regAdS2d-rho1}).

\section{\label{Discrete}The Discrete Branes}
In subsection \ref{irrAdS2d-rho2}, we give the details of our derivation of the $b^{-2}/2$ shift equation for the irregular discrete $AdS_2$ branes with $\rho=\rho_2$, that have been studied in \cite{SR1}. With the help of our new shift equation, we then discuss consistency of the solution proposed in \cite{SR1}. Afterwards, we proceed with the shift equations and their solutions for the cases irregular discrete at $\rho=\rho_1$, regular discrete at $\rho_2$ and regular discrete at $\rho_1$ in subsections \ref{irrAdS2d-rho1}, \ref{regAdS2d-rho2} and \ref{regAdS2d-rho1}. Apart from some tedious but yet important details, especially involving signs and complex phases, the calculations are as in subsection \ref{irrAdS2d-rho2}, which is why we go into slightly less detail in the subsequent cases.

\subsection{\label{irrAdS2d-rho2}Irregular \boldmath$AdS^{(d)}_2$ Branes - Gluing Map \boldmath$\rho_2$}
\subsubsection{Shift Equations for the Boundary One Point Amplitudes}
The gluing map is $\rho_2$. Choosing the irregular $u$-dependence, it restricts the one point function in the presence of boundary condition $\alpha$ to be of the form
\eq{\cor{\Theta_j(u\arrowvert z)}_{\alpha}=\abs{z-\bar{z}}^{-2h(j)}\abs{u+\bar{u}}^{2j}A_{\sigma}(j\arrowvert\alpha).}
The unknown function $A_{\sigma}(j\arrowvert\alpha)$ is the {\sl one point amplitude}. Note that it still depends on $\sigma:=\sgn(u+\bar{u})$. Its physical interpretation is that it describes the strength of coupling of a closed string with label $j$ to the brane labelled by $\alpha$. It is possible to obtain necessary conditions on $A_{\sigma}(j\arrowvert \alpha)$ by considering two point functions involving a degenerate field. This strategy has been pursued in \cite{PST}, \cite{GKS} and \cite{LOP} for degenerate field $\Theta_{1/2}$ (and in \cite{GKS} one case has also been treated using the degenerate field $\Theta_{b^{-2}/2}$, see Table \ref{T1}). However, only a few cases have been checked so far and refering once again to Table \ref{T1}, it becomes clear that lots of constraints (shift equations) remain to be computed.\\
Let us now illustrate the whole procedure for the irregular $AdS^{(d)}_2$ branes in case of a two point function involving the degenerate field $\Theta_{b^{-2}/2}$. This will lead us to a formerly unknown shift equation.\\
Using the Ward identities, the form of the two point function $G^{(2)}_{j,\alpha}(u_i\arrowvert z_i):=\cor{\Theta_{b^{-2}/2}(u_2\arrowvert z_2)\Theta_j(u_1\arrowvert z_1)}_{\alpha}$ can be partially fixed as
\spliteq{\label{irrB2pt-2}G^{(2)}_{j,\alpha}(u_1,u_2\arrowvert z_1,z_2)=&\abs{z_1-\bar{z}_1}^{2[h(b^{-2}/2)-h(j)]}\abs{z_1-\bar{z}_2}^{-4h(b^{-2}/2)}\cdot\\
&\cdot \abs{u_1+\bar{u}_1}^{2j-b^{-2}}\abs{u_1+\bar{u}_2}^{2b^{-2}}H^{(2)}_{j,\alpha}(u\arrowvert z),} 
where $H^{(2)}_{j,\alpha}(u\arrowvert z)$ is an unknown function of the crossing ratios 
\eq{\label{CrossRatio}z:=\frac{\abs{z_2-z_1}^2}{\abs{z_2-\bar{z}_1}^2} \hskip .5cm {\rm and} \hskip .5cm
u:=\frac{\abs{u_2-u_1}^2}{\abs{u_2+\bar{u}_1}^2}.}
Now, the standard Knizhnik-Zamolodchikov equations are used to deduce a partial differential equation for $H^{(2)}_{j,\alpha}(u\arrowvert z)$ (see Appendix \ref{KniZa}). Since one field operator is the degenerate field $\Theta_{b^{-2}/2}$, its space of solutions is finite dimensional, in fact it consists of three conformal blocks only, namely those for $j_{\pm}:=j\pm b^{-2}/2$ and $j_{\times}:=-j-1-b^{-2}/2$. Hence, the general solution reads
\eq{\label{H2}H^{(2)}_{j,\alpha}(u\arrowvert z)=\sum_{\epsilon=+,-,\times} a^{j}_{\epsilon}(\alpha){\cal F}^{s}_{j,\epsilon}(u\arrowvert z),}
where the conformal blocks ${\cal F}^{s}_{j,\epsilon}(u\arrowvert z)$ are given in Appendix \ref{KniZa}, and the $a^{j}_{\epsilon}(\alpha)$ are some still undetermined coefficients. They are fixed by using the bulk OPE of the two field operators on the L.H.S. and taking the appropriate limit $\abs{z_2-z_1}\rightarrow 0$ on the R.H.S. of (\ref{irrB2pt-2}). The $a^{j}_{\epsilon}(\alpha)$ will then generically turn out to be some product of bulk OPE coefficient times one point amplitude, which is why the $\alpha$-dependence occurs in the $a^{j}_{\epsilon}$-coefficients. We find (see Appendix \ref{LinComb-irr2-1} for details) that 
\eq{\label{a-coeffs}a^{j}_{\epsilon}(\alpha)=C_{\epsilon}(j)A_{\sigma}(j_{\epsilon}\arrowvert\alpha).}
where the $C_{\epsilon}(j)$ are bulk OPE coefficients, see Appendix \ref{OPE2} for their explicit expressions. The boundary two point function (\ref{irrB2pt-2}) is now determined exactly.\\
In order to get a shift equation, we take the limit $\Im(z_2)\rightarrow 0$. Upon doing this, we can use the bulk-boundary OPE on the L.H.S. of (\ref{irrB2pt-2}) to obtain (see the discussion in section \ref{Discrete-Continuous} together with appendix \ref{BulkBdryOPE2})\footnote{We take the signs $\sigma_1:=\sgn(u_1+\bar{u}_1)$ and $\sigma_2:=\sgn(u_2+\bar{u}_2)$ to be equal here: $\sigma_1=\sigma_2=:\sigma$. One could, of course, also take $\sigma_1=-\sigma_2=:\sigma$. This would lead to further constraints. Note, however, that for $\sigma_1=\sigma_2$, the crossing ratio $u$ lies in the interval $0<u<1$, whereas for $\sigma_1=-\sigma_2$ we have $u>1$. In the latter case one has to use different analytic continuations when taking $z\rightarrow 1-$. We have not worked that out.}:
\spliteq{G^{(2)}_{j,\alpha}(u_1,u_2\arrowvert z_1,z_2)\simeq & \abs{z_1-\bar{z_1}}^{-2h(j)}\abs{z_2-\bar{z_2}}^{-2h(b^{-2}/2)}\cdot\\
& \cdot\abs{u_1+\bar{u_1}}^{2j}\abs{u_2+\bar{u_2}}^{b^{-2}}A_{\sigma}(j\arrowvert\alpha)A_{\sigma}(b^{-2}/2\arrowvert\alpha).}
On the R.H.S. of (\ref{irrB2pt-2}), we just take our exact expression (involving the results (\ref{H2}) and (\ref{a-coeffs})) and perform the limit explicitly. For this, an analytic continuation is needed - see chapter \ref{CloseLook} and appendix \ref{irrBranes} for details. If we redefine the one point amplitude (see Appendix \ref{RefCnstr} for a motivation of {\sl this} particular redefinition)
\eq{\label{ReDef}f_{\sigma}(j)\equiv f_{\sigma}(j\arrowvert\alpha):=\nu_b^j\Gamma(1+b^2(2j+1))A_{\sigma}(j\arrowvert\alpha)} 
and equate the two expressions from L.H.S. and R.H.S., we arrive at our new additional shift equation for the irregular $AdS^{(d)}_2$ brane:
\eq{\label{irrAdS2d-rho2-Shift2}\ebrac{\Gamma(1+b^2)}^{-1}f_{\sigma}\brac{\frac{b^{-2}}{2}}f_{\sigma}(j)=f_{\sigma}\brac{j+\frac{b^{-2}}{2}}+\e^{-\i\pi b^{-2}}f_{\sigma}\brac{j-\frac{b^{-2}}{2}}.}
For completeness let us also write down the formerly known shift equation \cite{SR1} for the redefined one point amplitude (\ref{ReDef}). It is 
\spliteq{\label{irrAdS2d-rho2-Shift1}-\frac{1}{\pi}\Gamma(-b^2)\sin[2\pi b^2]\sin[\pi b^2(2j+1)]f_{\sigma}\brac{\frac{1}{2}}f_{\sigma}\brac{j}=&\\
=\sin[\pi b^2(2j+2)]f_{\sigma}\brac{j+\frac{1}{2}}-\sin[&\pi b^2 2j]f_{\sigma}\brac{j-\frac{1}{2}}.}

\subsubsection{Solving the Shift Equations}
The formlery known shift equation (\ref{irrAdS2d-rho2-Shift1}) is solved by \cite{SR1}
\eq{\label{irrAdS2d-rho2-Sol}f_{\sigma}(j\arrowvert m,n)=\frac{\i\pi\sigma \e^{\i\pi m}}{\Gamma(-b^2)\sin[\pi nb^2]}\e^{-\i\pi\sigma(m-\frac{1}{2})(2j+1)}\frac{\sin[\pi nb^2(2j+1)]}{\sin[\pi b^2(2j+1)]},}
with $n,m\in\mathbb{Z}$.\footnote{This is how the solution has been given in \cite{SR1}. In fact we only need $m\in\mathbb{Z}$ to satisfy equation (\ref{irrAdS2d-rho2-Shift1}).} Note that this also satisfies the reflection symmetry constraint (\ref{RefSymmCnstr-irr}). One checks however quite easily that it does not satisfy our new shift equation (\ref{irrAdS2d-rho2-Shift2}). Interestingly, the obstruction is precisely the term that stems from the ${\cal F}^{s}_{j,-}$ conformal block. Without it, the equation would be obeyed.

\subsection{\label{irrAdS2d-rho1}Irregular \boldmath$AdS_2^{(d)}$ Branes - Gluing Map \boldmath$\rho_1$}
\subsubsection{Shift Equations}
Choosing the irregular $u$-dependence, the gluing map $\rho_1$ restricts the one point function in the presence of boundary condition $\alpha$ to be of the form
\eq{\cor{\Theta_j(u\arrowvert z)}_{\alpha}=\abs{z-\bar{z}}^{-2h(j)}\abs{u-\bar{u}}^{2j}A_{\sigma}(j\arrowvert\alpha).}
Our ansatz for the boundary two point function with degenerate field $t/2$, $t=1,b^{-2}$ (fixing the $u_i$ and $z_i$ dependence up to a dependence on the crossing ratios) is
\spliteq{\label{irrB2pt-1}G^{(2)}_{j,t,\alpha}(u_1,u_2\arrowvert z_1,z_2)=\abs{z_1-\bar{z}_1}&^{2[h(t/2)-h(j)]}\abs{z_1-\bar{z}_2}^{-4h(t/2)}\cdot\\
&\cdot \abs{u_1-\bar{u}_1}^{2j-t}\abs{u_1-\bar{u}_2}^{2t}H^{(2)}_{j,t,\alpha}(u\arrowvert z),} 
with crossing ratios
\eq{z:=\frac{\abs{z_2-z_1}^2}{\abs{z_2-\bar{z}_1}^2} \hskip .5cm {\rm and} \hskip .5cm
u:=\frac{\abs{u_2-u_1}^2}{\abs{u_2-\bar{u}_1}^2}.}
The conformal blocks that solve the Knizhnik-Zamolodchikov equations turn out to be just the same ones as for gluing map $\rho_2$, so for $t=b^{-2}$ they are given by (\ref{ConfBlocks}) with parameters
\eq{\alpha=\beta=-b^{-2},\hskip .3cm \beta'=-2j-1-b^{-2},\hskip .3cm \gamma=-2j-b^{-2}}
and for $t=1$ see \cite{PST}.
Also, in both cases ($t=1,b^{-2}$), the expansion coefficients stay as before:
\eq{a^{j}_{\epsilon}(\alpha)=C_{\epsilon}(j)A_{\sigma}(j_{\epsilon}\arrowvert\alpha).}
Taking the limit $\Im(z_2)\rightarrow 0$, we obtain the same shift equations as for gluing map $\rho_2$, namely
\spliteq{\label{irrAdS2d-rho1-Shift1}-\frac{1}{\pi}\Gamma(-b^2)\sin[2\pi b^2]\sin[\pi b^2(2j+1)]f_{\sigma}\brac{\frac{1}{2}}f_{\sigma}\brac{j}=&\\
=\sin[\pi b^2(2j+2)]f_{\sigma}\brac{j+\frac{1}{2}}-\sin[&\pi b^2 2j]f_{\sigma}\brac{j-\frac{1}{2}}}
and
\eq{\label{irrAdS2d-rho1-Shift2}\ebrac{\Gamma(1+b^2)}^{-1}f_{\sigma}\brac{\frac{b^{-2}}{2}}f_{\sigma}(j)=f_{\sigma}\brac{j+\frac{b^{-2}}{2}}+\e^{-\i\pi b^{-2}}f_{\sigma}\brac{j-\frac{b^{-2}}{2}}.}
This means that the irregular discrete branes that arise from gluing maps $\rho_1$ and $\rho_2$ respectively, are indeed isomorphic. Compare to our remarks in the introduction and in chapter \ref{GlueConds}, where we explained that this is likely to happen. However, we do not find a solution that satsifies both factorization constraints (\ref{irrAdS2d-rho1-Shift1}), (\ref{irrAdS2d-rho1-Shift2}), as we have already explained in section \ref{irrAdS2d-rho2}.

\subsection{\label{regAdS2d-rho2}Regular \boldmath$AdS_2^{(d)}$ Branes - Gluing Map \boldmath$\rho_2$}
\subsubsection{Shift Equations}
This time choosing the regular $u$-dependence, the gluing map $\rho_2$ fixes the one point function as
\eq{\cor{\Theta_j(u\arrowvert z)}_{\alpha}=\brac{z-\bar{z}}^{-2h(j)}\brac{u+\bar{u}}^{2j}A_{\sigma}(j\arrowvert\alpha).}
The boundary two point function with degenerate field $t/2$, $t=1,b^{-2}$ is
\spliteq{\label{regB2pt-2}G^{(2)}_{j,t,\alpha}(u_1,u_2\arrowvert z_1,z_2)=\brac{z_1-\bar{z}_1}&^{-2h(j)}\brac{z_2-\bar{z}_2}^{-2h(t/2)}\cdot\\
&\cdot \brac{u_1+\bar{u}_1}^{2j}\brac{u_2+\bar{u}_2}^{t}H^{(2)}_{j,t,\alpha}(u\arrowvert z),} 
with crossing ratios
\eq{z:=\frac{\abs{z_1-z_2}^2}{\brac{z_1-\bar{z}_1}\brac{z_2-\bar{z}_2}} \hskip .5cm {\rm and} \hskip .5cm
u:=-\frac{\abs{u_1-u_2}^2}{\brac{u_1+\bar{u}_1}\brac{u_2+\bar{u}_2}}.}
They take values in $z\in(-\infty,0)$ (because $z_1$, $z_2$ are in the upper half plane), $u\in(-\infty,0)$ (if we take $\sigma_1=\sigma_2$). Solving the Knizhnik-Zamolodchikov equations for $t=1$ results in the following conformal blocks:
\spliteq{\label{ConfBlocks-1/2}{\cal F}^{s}_{j,+}(u\arrowvert z)&=z^{-b^{2}j}(1-z)^{-b^{2}j}\set{F(a,b;c\arrowvert z)-u\brac{\frac{b}{c}}F(a,b+1;c+1\arrowvert z)},\\
{\cal F}^{s}_{j,-}(u\arrowvert z)&=\left. z^{b^2(j+1)}(1-z)^{b^{2}j}\right\{uF(a-c,b-c+1;1-c\arrowvert z)-\\
&-\left. z\brac{\frac{a-c}{1-c}}F(a-c+1,b-c+1;2-c\arrowvert z)\right\}}
($F(a,b;c\arrowvert z)$ is the Gauss' hypergeometric function), with parameters 
\eq{a=-b^2(2j+2),\hskip .3cm b=-b^2(2j),\hskip .3cm c=-b^2(2j+1).}
The solution for $t=b^{-2}$ is provided by the conformal blocks (\ref{ConfBlocksReg}), which are very similar, but not identical, to the ones we encountered before. Also, the expansion coefficients have to be modified slightly in this case. For $t=1$, we have
\eq{a^{j,1/2}_{\epsilon}(\alpha)=\epsilon C^{1/2}_{\epsilon}(j)A_{\sigma}(j_{\epsilon}\arrowvert\alpha)}
and for $t=b^{-2}$
\spliteq{\label{a-coeffs-reg-2}a^{j,b^{-2}/2}_{+}(\alpha)&=\e^{-4\pi\i j}C^{b^{-2}/2}_{+}(j)A_{\sigma}(j_{+}\arrowvert\alpha),\\
a^{j,b^{-2}/2}_{-}(\alpha)&=\e^{4\pi\i j-\i\pi (\sigma-2)b^{-2}}C^{b^{-2}/2}_{-}(j)A_{\sigma}(j_{-}\arrowvert\alpha),\\
a^{j,b^{-2}/2}_{\times}(\alpha)&=\e^{-4\pi\i j-\i\pi(\sigma-2)(2j+1+b^{-2})}C^{b^{-2}/2}_{\times}(j)A_{\sigma}(j_{\times}\arrowvert\alpha).}
For details, especially about the complex phases, see appendix \ref{LinComb-reg2}. 
Taking the limit $\Im(z_2)\rightarrow 0$, we have this time that $z\rightarrow -\infty$. Therefore, we have to take different analytic continuations of the occuring Gauss hypergeometric and Appell functions than before. See appendix \ref{regBranes} for the details. We obtain the following shift equations
\spliteq{\label{regAdS2d-rho2-Shift1}&-\frac{1}{\pi}\Gamma(-b^2)\sin[2\pi b^2]\sin[\pi b^2(2j+1)]f_{\sigma}\brac{\frac{1}{2}}f_{\sigma}(j)=\\
&\hphantom{..}=\e^{\i\pi b^{2}j}\sin[\pi b^2(2j+2)]f_{\sigma}\brac{j+\frac{1}{2}}-\e^{-\i\pi b^{2}(j+1)}\sin[\pi b^2 2j]f_{\sigma}\brac{j-\frac{1}{2}}}
and
\spliteq{\label{regAdS2d-rho2-Shift2}\ebrac{\Gamma(1+b^2)}^{-1}f_{\sigma}\brac{\frac{b^{-2}}{2}}f_{\sigma}(j)&=\e^{-\i\pi 3j}f_{\sigma}\brac{j+\frac{b^{-2}}{2}}-\\
-\e^{\i\pi 3j}\e^{-\i\pi\sigma b^{-2}}f_{\sigma}\brac{j-\frac{b^{-2}}{2}}+\e^{-\i\pi 3j}&\e^{-\i\pi\sigma (2j+b^{-2})}f_{\sigma}\brac{-j-1-\frac{b^{-2}}{2}}.}

\subsubsection{Solving the Shift Equations}
To begin with, let us explain what we understand by the term "solution" in this situation. We have remarked earlier that the regular branes should only be considered for $j\in\frac{1}{2}\mathbb{Z}$ because of monodromy. Thus, one can ask what the meaning of a $b^{-2}/2$-shift equation shall be. We like to think of the one point function $f_{\sigma}(j)$ as beeing defined for general complex values of $j$. The solutions for the irregular branes had this property, but there was also no a priori restriction on the values of $j$. In view of the restriction $j\in\frac{1}{2}\mathbb{Z}$ that we are facing here, we like to think of a solution as an {\sl interpolating} solution, in the sense that it is defined for general $j\in\mathbb{C}$, but then keeping in mind that it actually only interests us on the half-integers. Assuming this, we can show that the $1/2$-shift equation (\ref{regAdS2d-rho2-Shift1}) together with the reflection symmetry constraint (\ref{RefSymmCnstr-reg}) does only admit $1$-periodic or $1$-antiperiodic solutions $f_{\sigma}(j)$, which is however incompatible with the $b^{-2}/2$-shift equation. Thus, a solution (in the sense explained above) does not exist. For details of the proof see appendix \ref{NoSolution}.

\subsection{\label{regAdS2d-rho1}Regular \boldmath$AdS_2^{(d)}$ Branes - Gluing Map \boldmath$\rho_1$}
\subsubsection{Shift Equations}
Again we choose the regular $u$-dependence, so that the gluing map $\rho_1$ fixes the one point function to be
\eq{\cor{\Theta_j(u\arrowvert z)}_{\alpha}=\brac{z-\bar{z}}^{-2h(j)}\brac{u-\bar{u}}^{2j}A_{\sigma}(j\arrowvert\alpha).}
The boundary two point function with degenerate field $t/2$, $t=1,b^{-2}$ is
\spliteq{\label{regB2pt-1}G^{(2)}_{j,t,\alpha}(u_1,u_2\arrowvert z_1,z_2)=\brac{z_1-\bar{z}_1}&^{-2h(j)}\brac{z_2-\bar{z}_2}^{-2h(t/2)}\cdot\\
&\cdot \brac{u_1-\bar{u}_1}^{2j}\brac{u_2-\bar{u}_2}^{t}H^{(2)}_{j,t,\alpha}(u\arrowvert z),} 
with crossing ratios
\eq{z:=\frac{\abs{z_1-z_2}^2}{\brac{z_1-\bar{z}_1}\brac{z_2-\bar{z}_2}} \hskip .5cm {\rm and} \hskip .5cm
u:=\frac{\abs{u_1-u_2}^2}{\brac{u_1-\bar{u}_1}\brac{u_2-\bar{u}_2}}.}
Note that again $z\in(-\infty,0)$, $z\in(-\infty,0)$. Solving the Knizhnik-Zamolodchikov equations for $t=1$ results in the same conformal blocks as for gluing map $\rho_2$, so they are given by equation (\ref{ConfBlocks-1/2}), again with parameters
\eq{a=-b^2(2j+2),\hskip .3cm b=-b^2(2j),\hskip .3cm c=-b^2(2j+1).}
The solution for $t=b^{-2}$ yields the conformal blocks (\ref{ConfBlocks}), again with parameters
\eq{\alpha=-2j,\hskip .3cm \beta=-b^{-2},\hskip .3cm \beta'=-2j-1-b^{-2},\hskip .3cm \gamma=-2j-b^{-2},}
just like for $\rho_2$.
The expansion coefficients for $t=1$ are not altered here. But those for $t=b^{-2}$ again acquire complex phases:
\spliteq{\label{a-coeffs-reg-1}a^{j,b^{-2}/2}_{+}(\alpha)&=\e^{-4\pi\i j}C^{b^{-2}/2}_{+}(j)A_{\sigma}(j_{+}\arrowvert\alpha),\\
a^{j,b^{-2}/2}_{-}(\alpha)&=\e^{4\pi\i j+\i\pi (\sigma+3)b^{-2}}C^{b^{-2}/2}_{-}(j)A_{\sigma}(j_{-}\arrowvert\alpha),\\
a^{j,b^{-2}/2}_{\times}(\alpha)&=\e^{-4\pi\i j+\i\pi(\sigma+3)(2j+b^{-2})}C^{b^{-2}/2}_{\times}(j)A_{\sigma}(j_{\times}\arrowvert\alpha).}
In the limit $\Im(z_2)\rightarrow 0$, the same comments as in chapter \ref{regAdS2d-rho2} for $\rho_2$ apply. The shift equations that we produce read
\spliteq{\label{regAdS2d-rho1-Shift1}&-\frac{1}{\pi}\Gamma(-b^2)\sin[2\pi b^2]\sin[\pi b^2(2j+1)]f_{\sigma}\brac{\frac{1}{2}}f_{\sigma}(j)=\\
&\hphantom{..}=\e^{\i\pi b^{2}j}\sin[\pi b^2(2j+2)]f_{\sigma}\brac{j+\frac{1}{2}}+\e^{-\i\pi b^{2}(j+1)}\sin[\pi b^2 2j]f_{\sigma}\brac{j-\frac{1}{2}}}
and
\spliteq{\label{regAdS2d-rho1-Shift2}\ebrac{\Gamma(1+b^2)}^{-1}f_{\sigma}\brac{\frac{b^{-2}}{2}}f_{\sigma}(j)=\e^{-\i\pi 3j}&f_{\sigma}\brac{j+\frac{b^{-2}}{2}}-\\
-\e^{\i\pi (3j+b^{-2})}\e^{\i\pi\sigma b^{-2}}f_{\sigma}\brac{j-\frac{b^{-2}}{2}}-\e^{-\i\pi (j-b^{-2})}&\e^{\i\pi\sigma (2j+b^{-2})}f_{\sigma}\brac{-j-1-\frac{b^{-2}}{2}}.}
Note that the $1/2$-shift equation (\ref{regAdS2d-rho1-Shift1}) differs from the one for $\rho_2$, (\ref{regAdS2d-rho2-Shift1}), only in the sign between the two terms on the R.H.S. It is this little detail that allows for a more general solution than before in the case of $\rho_2$.

\subsubsection{Solving the Shift Equations}
Regarding the properties we like to assume of a solution, the same comments as in section \ref{regAdS2d-rho2} apply. We take a first step by solving the $1/2$-shift equation (\ref{regAdS2d-rho1-Shift1}) together with the reflection symmetry constraint (\ref{RefSymmCnstr-reg}). We find the following two parameter solution
\eq{\label{regAdS2d-rho1-Sol}f_{\sigma}(j\arrowvert m,n)=-\frac{\pi\e^{\i\pi\frac{b^2}{4}}\e^{\i\pi m}}{\Gamma(-b^2)\sin[\pi nb^2]}\e^{-\i\pi\sigma (2j+1)m}\e^{-\i\pi\frac{b^2}{4}(2j+1)^2}\frac{\sin[\pi nb^2 (2j+1)]}{\sin[\pi b^2 (2j+1)]},}
with $m\in\mathbb{Z}$  and up to now no restrictions on the parameter $n$. Comparing to the solution for the irregular branes (\ref{irrAdS2d-rho2-Sol}), it differs essentially in the term $\propto\exp(-\i\pi\frac{b^2}{4}(2j+1)^2)$. Note that this is an additional quantum deformation, since for $b^2\rightarrow 0$ (corresponding to $k\rightarrow\infty$) this term goes to one. In the classical limit, our solution behaves like $\propto\exp\ebrac{-\sigma (2j+1)r}$, what is the expected behaviour for an $AdS_2$ brane (see \cite{PST}). Furthermore, note that $\e^{\i\pi\sigma (2j+1)m}$ does not depend on $\sigma$ when $j\in\frac{1}{2}\mathbb{Z}$ (remember that the regular branes should only be considered for that case). So, on the half-integers this solution is actually independent of $\sigma$.\\
Inserting (\ref{regAdS2d-rho1-Sol}) into the second shift equation (\ref{regAdS2d-rho1-Shift2}), we find, very remarkably, that is obeyed provided that $n\in\mathbb{Z}$ and $j\in\frac{1}{2}\mathbb{Z}$ is used in the last step. Thus, in (\ref{regAdS2d-rho1-Sol}) with $m,n\in\mathbb{Z}$ and $j\in\frac{1}{2}\mathbb{Z}$, we have given a fully consistent solution to both factorization constraints.

\section{\label{Continuous}The Continuous Branes}
In this chapter we assemble our results (shift equations and solutions) concerning the continuous branes. The two point functions are always determined as shown in the corresponding sections of chapter \ref{Discrete} and thus, we do not write them down here again, but merely state our results. Recall from section \ref{Discrete-Continuous} that, instead of $C_{\sigma}=A_{\sigma}$, we now encounter the residua $\tilde{c}_{\sigma}$ on the L.H.S. The $1/2$-shift equations for the irregular continuous branes with gluing maps $\rho_2$ and $\rho_1$ have already been discussed in \cite{PST} and \cite{LOP}.

\subsection{\label{irrAdS2c-rho2-1}Irregular \boldmath$AdS_2^{(c)}$ Branes - Gluing Maps \boldmath$\rho_1$, \boldmath$\rho_2$}
As before in the discrete case, we discover that the irregular continuous branes are isomorphic for gluing maps $\rho_1$, $\rho_2$. The shift equations are
\spliteq{\label{irrAdS2c-rho2-1-Shift1}\sqrt{\nu_b}\frac{\Gamma(-b^2)}{\Gamma(-2b^2)}\tilde{c}_{\sigma}(1/2,0\arrowvert\alpha)\sin[\pi b^2(2j+1)]f_{\sigma}\brac{j}=&\\
=\sin[\pi b^2(2j+2)]f_{\sigma}\brac{j+\frac{1}{2}}-&\sin[\pi b^2 2j]f_{\sigma}\brac{j-\frac{1}{2}}}
and
\spliteq{\label{irrAdS2c-rho2-1-Shift2}(1+b^2)\nu_b^{\frac{b^{-2}}{2}}\tilde{c}_{\sigma}(b^{-2}/2,0\arrowvert\alpha)f_{\sigma}(j)&=f_{\sigma}\brac{j+\frac{b^{-2}}{2}}+\e^{-\i\pi b^{-2}}f_{\sigma}\brac{j-\frac{b^{-2}}{2}}.}
In \cite{PST} and \cite{LOP}, the following solution to the $1/2$-shift equation (\ref{irrAdS2c-rho2-1-Shift1}) and the reflection symmetry constraint (\ref{RefSymmCnstr-irr}) has been proposed
\eq{\label{irrAdS2c-rho2-1-Sol}f_{\sigma}(j\arrowvert\alpha)=-\frac{\pi A_b}{\sqrt{\nu_b}}\frac{\e^{-\alpha(2j+1)\sigma}}{\sin[\pi b^2(2j+1)]}.}
To obtain this solution, it was used that 
\eq{\label{irrAdS2c-rho2-1-c-1}\tilde{c}_{\sigma}(1/2,0\arrowvert\alpha)=-\frac{\sigma}{\sqrt{\nu_b}}\frac{\Gamma(-2b^2)}{\Gamma(-b^2)}2\sinh(\alpha).}
Plugging the solution (\ref{irrAdS2c-rho2-1-Sol}) into the $b^{-2}/2$-shift equation (\ref{irrAdS2c-rho2-1-Shift2}), we can infer an expression for the unknown $\tilde{c}_{\sigma}(b^{-2}/2,0\arrowvert\alpha)$:
\eq{\label{irrAdS2c-rho2-1-c-2}\tilde{c}_{\sigma}(b^{-2}/2,0\arrowvert\alpha)=-\frac{\e^{-\i\pi b^{-2}/2}}{\nu_b^{b^{-2}/2}(1+b^2)}2\cosh\ebrac{\brac{\alpha\sigma-\i\frac{\pi}{2}}b^{-2}}.}
Hence, the known irregular continuous $AdS_2$ branes are fully consistent with both factorization constraints. Note that for $b^{-2}=b^2=1$, the bulk-boundary OPE coefficients (\ref{irrAdS2c-rho2-1-c-1}), (\ref{irrAdS2c-rho2-1-c-2}) coincide, as do the two shift equations (\ref{irrAdS2c-rho2-1-Shift1}) and (\ref{irrAdS2c-rho2-1-Shift2}). 

\subsection{\label{regAdS2c}Regular \boldmath$AdS_2^{(c)}$ Branes - Gluing Maps \boldmath$\rho_1$, \boldmath$\rho_2$}
We have just seen that knowledge about the occuring coefficients $\tilde{c}_{\sigma}(1/2,0\arrowvert\alpha)$ and $\tilde{c}_{\sigma}(b^{-2}/2,0\arrowvert\alpha)$ is needed to decide whether the continuous branes are consistent or not. In \cite{PST}, $\tilde{c}_{\sigma}(1/2,0\arrowvert\alpha)$ has been given for irregular branes. However, let us point out that we cannot expect the corresponding coefficients in the regular case $\tilde{c}^{(reg)}_{\sigma}(1/2,0\arrowvert\alpha)$, $\tilde{c}^{(reg)}_{\sigma}(b^{-2}/2,0\arrowvert\alpha)$ to coincide with the irregular ones. That is why we will put a superscript from now on. That we need to distinguish the coefficients between regular and irregular case is already indicated in the discrete branes: There, the bulk-boundary OPE coefficient $C_{\sigma}(1/2,0|m,n)$ is identified with the one point amplitude $A_{\sigma}(1/2|m,n)$. Now, comparing the solutions for the one point amplitude (\ref{irrAdS2d-rho2-Sol}), (\ref{regAdS2d-rho1-Sol}) at $j=1/2$ reveals that the bulk-boundary OPE coefficients do not coincide, but are related as
\eq{\label{C-Relation}C_{\sigma}^{(reg)}(1/2,0|m,n)=-\i\sigma \e^{-\i\pi\frac{3}{4}b^2}C_{\sigma}^{(irr)}(1/2,0|m,n).}
Note that we need to be careful here, because the irregular one point amplitude does actually not satisfy the second factorization constraint, and it is thus questionable if $C_{\sigma}^{(irr)}(1/2,0|m,n)$ on the RHS is sensible and correct. Yet, for the moment we take (\ref{C-Relation}) as a hint that the bulk-boundary coefficients (and with them the residua occuring for continuous branes) are not identical for regular and irregular branes, but closely related.\\
Therefore, when studying solutions to the shift equations for regular continuous $AdS_2$ branes in the next sections, our approach will be not to make any a priori assumptions about $\tilde{c}^{(reg)}_{\sigma}(1/2,0\arrowvert\alpha)$ and $\tilde{c}^{(reg)}_{\sigma}(b^{-2}/2,0\arrowvert\alpha)$. Instead, our guiding principle will be a certain ansatz for the form of the (redefined) one point amplitude. By inserting this ansatz into the shift equations, we will infer expressions for $\tilde{c}^{(reg)}_{\sigma}(1/2,0\arrowvert\alpha)$ and $\tilde{c}^{(reg)}_{\sigma}(b^{-2}/2,0\arrowvert\alpha)$ that we then discuss. So, let us explain the general ansatz for the regular one point amplitudes that we are going to use: In order to be $AdS_2$ branes, the solutions should behave like $\propto \exp[-\sigma(2j+1)r]$ in the classical limit $b^2\rightarrow 0$ (compare \cite{PST} and our remarks in section \ref{regAdS2d-rho1}). Secondly, from our experience in the discrete case (section \ref{regAdS2d-rho1}), we expect the additional quantum deformation $\exp[-\i\pi\frac{b^2}{4}(2j+1)]$ to occur. (You might remember from the discrete branes that, technically, this is the term that cancels the $\exp{[\i\pi b^2 j]}$ and $\exp[-\i\pi b^2 (j+1)]$ factors on the RHS of the $1/2$-shift equation; these factors are of course also present here, in the continuous case). Thirdly, just like in the irregular continuous solution, we also expect the deformation $\sin^{-1}[\pi b^2 (2j+1)]$ to be present. Along with it, an additional factor of $\sigma$ has to be included in order to get the parity of the solution right (recall that, from the reflection symmetry, regular branes must have parity opposite to the irregular ones; see appendix \ref{RefCnstr}). Putting all this together, the most natural ansatz for regular continuous branes is
\eq{\label{regAdS2c-Ansatz}f_{\sigma}^{(reg)}(j|\alpha)=A_{b}^{(reg)}\sigma \e^{-\i\pi\frac{b^2}{4}(2j+1)^2}\frac{\e^{-\sigma (2j+1)\alpha}}{\sin[\pi b^2 (2j+1)]},}
with an arbitrary, but only $b$-dependent constant $A_{b}^{(reg)}$. Just as above, in the irregular continuous case, it cannot be fixed, because the continuous shift equations are always linear in the one point amplitude. (\ref{regAdS2c-Ansatz}) is the form of solution we are going to plug into the shift equations for gluing maps $\rho_2$, $\rho_1$ in the next two subsections.

\subsubsection{\label{regAdS2c-rho2}Gluing Map \boldmath$\rho_2$}
We have the following shift equations
\spliteq{\label{regAdS2c-rho2-Shift1}&\sqrt{\nu_b}\frac{\Gamma(-b^2)}{\Gamma(-2b^2)}\tilde{c}(1/2,0\arrowvert\alpha)\sin[\pi b^2(2j+1)]f_{\sigma}\brac{j}=\\
&\hphantom{..}=\e^{\i\pi b^{2}j}\sin[\pi b^2(2j+2)]f_{\sigma}\brac{j+\frac{1}{2}}-\e^{-\i\pi b^{2}(j+1)}\sin[\pi b^2 2j]f_{\sigma}\brac{j-\frac{1}{2}}}
and
\spliteq{\label{regAdS2c-rho2-Shift2}(1+b^2)\nu_b^{\frac{b^{-2}}{2}}\tilde{c}(b^{-2}/2,0\arrowvert\alpha)f_{\sigma}(j)&=\e^{-\i\pi 3j}f_{\sigma}\brac{j+\frac{b^{-2}}{2}}-\\
-\e^{\i\pi 3j}\e^{-\i\pi\sigma b^{-2}}f_{\sigma}\brac{j-\frac{b^{-2}}{2}}+\e^{-\i\pi 3j}&\e^{-\i\pi\sigma (2j+b^{-2})}f_{\sigma}\brac{-j-1-\frac{b^{-2}}{2}}.}
From the $1/2$-shift equation, our ansatz yields
\eq{\label{regAdS2c-rho2-c-1}\tilde{c}_{\sigma}^{(reg)}(1/2,0|\alpha)=-\sigma\frac{\e^{-\i\pi\frac{3}{4}b^2}}{\sqrt{\nu_b}}\frac{\Gamma(-2b^2)}{\Gamma(-b^2)}2\sinh(\alpha).} 
Comparing to the irregular case we have
\eq{\tilde{c}_{\sigma}^{(reg)}(1/2,0|\alpha)=\e^{-\i\pi\frac{3}{4}b^2}\tilde{c}_{\sigma}^{(irr)}(1/2,0|\alpha),}
which fits our expectations. However, plugging our ansatz into the $b^{-2}/2$-shift equation, we obtain
\spliteq{-\i\e^{\i\pi\frac{b^{-2}}{4}}(1+b^2)\nu_b^{b^{-2}/2}\tilde{c}_{\sigma}(b^{-2}/2,0|\alpha)&=\e^{-\sigma\alpha b^{-2}}\e^{-4\pi\i j}+\\
+\e^{-\i\pi\sigma b^{-2}}\e^{\sigma\alpha b^{-2}}\e^{4\pi\i j}-\e^{2\sigma (2j+1)\alpha}&\e^{-\i\pi\sigma (2j+b^{-2})}\e^{-\sigma\alpha b^{-2}}\e^{-4\pi\i j}.}
The RHS is not independent of $j$, even if we take $j\in\frac{1}{2}\mathbb{Z}$. Therefore, further restrictions on $\alpha$ have to be made. For $j\in\frac{1}{2}\mathbb{Z}$, independence of $j$ is achieved if 
\eq{\alpha\in\i\pi\brac{\mathbb{Z}+\frac{1}{2}}.}
If this is the case, we can read off that 
\eq{\tilde{c}^{(reg)}_{\sigma}(b^{-2}/2,0|\alpha)=  \i\frac{\e^{-\i\pi\frac{b^{-2}}{4}}}{(1+b^2)\nu_b^{b^{-2}/2}}\brac{\e^{-\sigma\alpha b^{-2}}+\e^{-\i\pi\sigma b^{-2}}\cdot 2\cosh(\alpha b^{-2})}.}
Comparing to the irregular expression(\ref{irrAdS2c-rho2-1-c-2}), this does not seem very natural. Together with the quite peculiar restriction on $\alpha$, this "solution" is not very attractive.

\subsubsection{\label{regAdS2c-rho1}Gluing Map \boldmath$\rho_1$}
For the shift equations, we obtain
\spliteq{\label{regAdS2c-rho1-Shift1}&\sqrt{\nu_b}\frac{\Gamma(-b^2)}{\Gamma(-2b^2)}\tilde{c}(1/2,0\arrowvert\alpha)\sin[\pi b^2(2j+1)]f_{\sigma}\brac{j}=\\
&\hphantom{..}=\e^{\i\pi b^{2}j}\sin[\pi b^2(2j+2)]f_{\sigma}\brac{j+\frac{1}{2}}+\e^{-\i\pi b^{2}(j+1)}\sin[\pi b^2 2j]f_{\sigma}\brac{j-\frac{1}{2}}}
and
\spliteq{\label{regAdS2c-rho1-Shift2}(1+b^2)\nu_b^{\frac{b^{-2}}{2}}\tilde{c}(b^{-2}/2,0\arrowvert\alpha)f_{\sigma}(j)=\e^{-\i\pi 3j}&f_{\sigma}\brac{j+\frac{b^{-2}}{2}}-\\
-\e^{\i\pi (3j+b^{-2})}\e^{\i\pi\sigma b^{-2}}f_{\sigma}\brac{j-\frac{b^{-2}}{2}}-\e^{-\i\pi (j-b^{-2})}&\e^{\i\pi\sigma (2j+b^{-2})}f_{\sigma}\brac{-j-1-\frac{b^{-2}}{2}}.}
Using our ansatz in the $1/2$-shift equation results in
\eq{\label{regAdS2c-rho1-c-1}\tilde{c}_{\sigma}^{(reg)}(1/2,0|\alpha)=\frac{\e^{-\i\pi\frac{3}{4}b^2}}{\sqrt{\nu_b}}\frac{\Gamma(-2b^2)}{\Gamma(-b^2)}2\cosh(\alpha).}
The $b^{-2}/2$-shift equation, however, makes the same problems as above. This time, it turns out that we need the additional restriction
\eq{\alpha\in\i\pi\mathbb{Z}}
(taking $j\in\frac{1}{2}\mathbb{Z}$) which finally results in
\spliteq{\tilde{c}^{(reg)}_{\sigma}(b^{-2}/2,0|\alpha)=\i\frac{\e^{-\i\pi\frac{b^{-2}}{4}}}{(1+b^2)\nu_b^{b^{-2}/2}}&\left(\e^{-\sigma\alpha b^{-2}}+\e^{\i\pi b^{-2}}\e^{\sigma\alpha b^{-2}}\right.+\\
&\hphantom{(}+\left.\e^{\i\pi\sigma b^{-2}}\e^{\i\pi b^{-2}}\e^{-\sigma\alpha b^{-2}}\right),}
both again not very natural results.

\subsubsection{One last remark about the regular \boldmath$AdS_2^{(c)}$ branes with gluing map \boldmath$\rho_1$}
Compare the discrete one point functions (\ref{irrAdS2d-rho2-Sol}) and (\ref{regAdS2d-rho1-Sol}): They essentially 
differ in parity, the additional quantum deformation $\exp[-\i\pi\frac{b^2}{4}(2j+1)]$ and a shifted parameter
\eq{m^{(reg)}=m^{(irr)}+\frac{1}{2}.}
We could try to incorporate such a shift for the continuous branes as well. It seems attractive to use
\eq{\alpha^{(reg)}=\alpha^{(irr)}+\i\pi\frac{1}{2}.}
With this shift, the $1/2$-shift equation (\ref{regAdS2c-rho1-Shift1}) is satisfied with
\eq{\tilde{c}_{\sigma}^{(reg)}(1/2,0|\alpha)=-\i\sigma\frac{\e^{-\i\pi\frac{3}{4}b^2}}{\sqrt{\nu_b}}\frac{\Gamma(-2b^2)}{\Gamma(-b^2)}2\sinh(\alpha).} 
Comparing to the irregular case, this is precisely
\eq{\tilde{c}_{\sigma}^{(reg)}(1/2,0|\alpha)=-\i\sigma \e^{-\i\pi\frac{3}{4}b^2}\tilde{c}_{\sigma}^{(irr)}(1/2,0|\alpha),}
what coincides with our observation form the discrete case (\ref{C-Relation}). Yet, let this shift be as attractive as it is, the $b^{-2}/2$-shift equation (\ref{regAdS2c-rho1-Shift2}) still suffers from the same problems as above. From the unnaturalness of the restrictions required to make both factorization constraints work, we would conjecture at this point that the regular continuous branes are actually not consistent.

\section{{\label{CloseLook}}A Closer Look at the Irregular Boundary Two Point Function}
In this section, we want to examine the process of analytic continuation more carefully. We are focussing on the irregular branes. The procedure for the regular ones is completely analogous to the treatment presented here. Let us take the ${\cal F}^{s}_{j,-}$ block as prototype example, since it shows all features that can be important:
\spliteq{{\cal F}^{s}_{j,-}(u\arrowvert z)&=z^{-j}(1-z)^{-b^{-2}/2}u^{-\beta}z^{1+\beta-\gamma}\cdot\\
&\cdot F_1\left(1+\beta+\beta'-\gamma,\beta,1+\alpha-\gamma;2+\beta-\gamma\left\arrowvert\frac{z}{u};z\right)\right..}
The conformal blocks we are using (\ref{ConfBlocks}) are well defined in the region $z<u<1$ (recall that the crossing ratios $u$ and $z$, as given in equation (\ref{CrossRatio}), are both real and $u,z\geq 0$). Since we like to maintain $u<1$, which corresponds to equal signs $\sigma_1=\sigma_2$ (recall that the one point amplitude $A_{\sigma}(j\arrowvert\alpha)$ depends on $\sigma=\sgn(u+\bar{u})$) and the factorization limit requires $z\rightarrow 1$, we necessarily need to continue to a patch where $z>u$, $u<1$, $z\approx 1$. We cannot use the standard analytic continuation of Appell's function $F_1$ as given in \cite{Exton}, because some coefficients turn out to become infinite in these formulae. This is due to the following relation between the parameters:
\eq{1+\beta'-\gamma=0.}
The invalidation of the continuation formulae in \cite{Exton} can be traced back to a special (logarithmic) case in the continuation of Gauss' hypergeometric function, when expanding $F_1$ appropriately (see appendix \ref{Hypergeo}). We will see this in detail shortly. In order to continue ${\cal F}^{s}_{j,-}$, the first step is to expand the occuring $F_1(\dots|\frac{z}{u};z)$ in powers of the first variable $\frac{z}{u}$ (see appendix \ref{AppellHorn})
\spliteq{F_1\left(\beta,\beta,1+\beta-\gamma;2+\beta-\gamma\left|\frac{z}{u};z\right)\right.=&\\
=\sum_{n=0}^{\infty}\frac{(\beta)_n(\beta_n)}{(2+\beta-\gamma)_n}F(\beta+n,1+\beta-\gamma&;2+\beta-\gamma+n|z)\frac{(z/u)^n}{n!}}
and then use a standard analytic continuation (as found e.g. in \cite{Bateman}) of Gauss' hypergeometric function $F(\dots|z)$ in order to expand it in terms of $(1-z)$. As can be seen from the parameters, this is a generic case. Furthermore, since $0\leq z<1$, also $0<(1-z)\leq 1$, meaning that no branch cuts are met and convergence in the domain needed is ensured. Two different terms arise from this continuation:
\spliteq{F_1\left(\beta,\beta,1+\beta-\gamma;2+\beta-\gamma\left|\frac{z}{u};z\right)\right.=:&\frac{\Gamma(2+\beta-\gamma)\Gamma(1-\beta)}{\Gamma(2-\gamma)}I\brac{\frac{z}{u};1-z}+\\
+\frac{\Gamma(2+\beta-\gamma)\Gamma(\beta-1)}{\Gamma(\beta)\Gamma(1+\beta-\gamma)}&(1-z)^{1-\beta}II\brac{\frac{z}{u};1-z}.}
Let us focus on the first one. After some minor manipulations it reads:
\eq{I\brac{\frac{z}{u};1-z}=\sum_{n=0}^{\infty}\frac{(\beta_n)(\beta)_n}{(1)_n}F(\beta+n,1+\beta-\gamma;\beta|1-z)\frac{(z/u)^n}{n!}.}
Now, we expand the hypergeometric function in powers of $(1-z)$ to yield a double expansion. Afterwards, the whole expression can be resummed and written as a single expansion again, but this time in powers of $(1-z)$ only:
\eq{I\brac{\frac{z}{u};1-z}=\sum_{m=0}^{\infty}(1+\beta-\gamma)_m F\left(\beta,\beta+m;1\left|\frac{z}{u}\right)\right.\frac{(1-z)^m}{m!}\,.}
In order to reach the desired patch, the "inner" hypergeometric function must now be continued to yield an expansion in the variable $\frac{u}{z}$. This, however, is no longer a generic case, but a logarithmic one. It is precisely where the formula for the full Appell function $F_1$ in \cite{Exton} breaks down. Nevertheless, we can do it right here. The appropriate continuation formula for the Gauss function is found in \cite{Bateman}, for example. We have also included it in appendix \ref{Hypergeo}. After its use, the resulting series it not easily resummend again to yield some familiar functions. But since we are taking the limit $z\rightarrow 1$ anyway, we can isolate the leading term in $(1-z)$, which is just the term with $m=0$ in the above expansion. Thus, for $z\rightarrow 1$, the result is
\eq{\label{I}I\simeq\frac{\e^{\i\pi\beta}u^{\beta}}{\Gamma(1-\beta)\Gamma(\beta)}\sum_{n=0}^{\infty}\frac{(\beta)_n(\beta)_n}{(1)_n}\frac{u^n}{n!}\ebrac{-\log(u)+h_n(\beta)-i\pi}\set{1+{\cal O}(1-z)}.} 
Note that the $u$ dependence $\propto\log(u)$ looks rather unfamiliar, but is actually nothing to worry about: The correct expansion variable for the OPE needed here is actually $(1-u)$ and $-\log(u)=-\log(1-(1-u))=(1-u)\set{1+{\cal O}(1-u)}$. Together with the prefactor $z^{-j}(1-z)^{-b^{-2}/2}u^{-\beta}z^{1+\beta-\gamma}$, which belongs to the definition of ${\cal F}^{s}_{j,-}$, this term has the correct asymptotics corresponding to propagation of the modes $b^{-2}$ and $-b^{-2}-1$ (see appendices \ref{irrBranes} and \ref{BulkBdryOPE2}). It does however not contribute to the propagation of the identity and consequently does not enter the factorization constraint.\\
There is still one more comment to make about the above continuation of Gauss' hypergeometric function from $\eta:=\frac{z}{u}$ to $\frac{u}{z}=\frac{1}{\eta}$. The Gauss function $F(a,b;c|\eta)$ has a branch cut along the line 
$\eta\in\mathbb{R}_{>1}$. Continuation formulae are invalidated if $\eta$ or its transformed counterpart take values in this line. This is, however, precisely the situation we need to handle. Let us explain how it can be done: Given the Gauss function $F(a,b;c|\eta)$, with $\eta\in\mathbb{R}$, $\eta>1$, let $\eta\mapsto\eta\e^{-\i\epsilon}$ ($\epsilon >0$). Gauss' hypergeometric function is continuous from below in $\eta$ (but not from above for $\eta>1$, so there is no choice involved here), i.e.
\eq{\label{EpsilonPresc}F(a,b;c|\eta)=\lim_{\epsilon\rightarrow 0+}F(a,b;c|\eta\e^{-\i\epsilon}).}
We therefore take occuring phases to be in $(-2\pi,0]$. In particular, this means $(-)=\e^{-\i\pi}$. On the RHS of (\ref{EpsilonPresc}), an analytic continuation formula (see \cite{Bateman}) can now be used. In the end, the epsilon is removed by taking it to be zero. This procedure automatically selects the correct phases. In practice, all we need to do is keep the phase prescription in mind and write everything without the epsilon. In the logarithmic case $b=a+m$, $m\in\mathbb{Z}_{\geq 0}$, the continuation is
\spliteq{F\left(a,a+m;c\left|\eta \right)\right.&=\frac{\Gamma(c)\brac{-\eta}^{-a-m}}{\Gamma(a+m)\Gamma(c-a)}\cdot\\
\cdot\sum_{n=0}^{\infty}\frac{(a)_{n+m}(1-c+a)_{n+m}}{n!(n+m)!}&\brac{\frac{1}{\eta}}^{n}\ebrac{\log(-\eta)+h_n(a,c,m)}+\\
+\frac{\Gamma(c)\brac{-\eta}^{-a}}{\Gamma(a+m)}\cdot&\sum_{n=0}^{m-1}\frac{\Gamma(m-n)(a)_n}{\Gamma(c-a-n)n!}\brac{\frac{1}{\eta}}^{n}.}
With our phase prescription, the logarithm becomes $\log(-\eta)=-\log(\frac{1}{\eta})-\i\pi$ and $(-\eta)^{-a}=\e^{\i\pi a}(\frac{1}{\eta})^a$. This is how the phase $\e^{\i\pi\beta}$ and the $-\i\pi$ in (\ref{I}) arise.\\
Let us now turn to the second term in the continuation of $F_1(\dots|\frac{z}{u};z)$:
\eq{II\brac{\frac{z}{u};1-z}=\sum_{n=0}^{\infty}(\beta)_n F(2-\gamma,1+n;2-\beta|1-z)\frac{(z/u)^n}{n!}.}
Expanding and resumming as above, this can equally be written as
\eq{II\brac{\frac{z}{u};1-z}=\sum_{m=0}^{\infty}\frac{(2-\gamma)_m (1)_m}{(2-\beta)_m}F\brac{\beta,1+m;1\left|\frac{z}{u}\right.}\frac{(1-z)^m}{m!}\,.}
This time, the continuation from $\frac{z}{u}$ to $\frac{u}{z}$ follows a generic case. The phase prescription is exactly as above. In the end, as $z\rightarrow 1$, we obtain
\eq{II\simeq\e^{\i\pi\beta}u^{\beta}F(\beta,\beta;\beta|u)\set{1+{\cal O}(1-z)}\,.}
Together with the overall prefactor $z^{-j}(1-z)^{1+b^{-2}/2}u^{-\beta}z^{1+\beta-\gamma}$ (coming from the definition of ${\cal F}^{s}_{j,-}$ together with the $(1-z)^{1-\beta}$ from the first continuation) and using that
\eq{F(\beta,\beta;\beta|u)=(1-u)^{-\beta}\,,}
this shows precisely the asymptotic behaviour of the propagating identity. This term therefore enters the factorization constraint.\\
The ${\cal F}^{s}_{j,\times}$ block 
\spliteq{{\cal F}^{s}_{j,\times}(u\arrowvert z)&=z^{-j}(1-z)^{-b^{-2}/2}u^{1-\gamma}\cdot\\
&\cdot G_2\left(\beta',1+\alpha-\gamma;1+\beta-\gamma,\gamma-1\left\arrowvert -\frac{z}{u};u\right.\right).}
can be treated along similar lines. Here, we first continue the second variable $u$ to $(1-u)$. As this turns out to be a generic case, the resumming works out again and we can then continue in the firrst variable from $\frac{z}{u}$ to $\frac{u}{z}$. This is again generic. The overall result does not contain a term corresponding to the identity propagating and hence no contribution to the shift equation is generated here.\\
Finally, the ${\cal F}^{s}_{j,+}$ block
\eq{{\cal F}^{s}_{j,+}(u\arrowvert z)=z^{-j}(1-z)^{-b^{-2}/2}F_1(\alpha,\beta,\beta';\gamma\arrowvert u;z)}
is easily continued using standard formulae of e.g. \cite{Exton}. We obtain
\spliteq{{\cal F}^{s}_{j,+}&\simeq (1-z)^{1+b^{-2}/2}(1-u)^{b^{-2}}\frac{\Gamma(\gamma)\Gamma(\alpha+\beta'-\gamma)}{\Gamma(\alpha)\Gamma(\beta')}\cdot\ebrac{1+{\cal O}(1-z)}+\\
&+(1-z)^{b^{-2}/2}\frac{\Gamma(\gamma)\Gamma(\gamma-\alpha-\beta')}{\Gamma(\gamma-\alpha)\Gamma(\gamma-\beta')}F(\alpha,\beta;\gamma-\beta'\arrowvert u)\cdot\ebrac{1+{\cal O}(1-z)}\,.}
The first summand gives the identity contribution and enters the shift equation.\\
While the original two point function, using the conformal blocks (\ref{ConfBlocks}), was defined in the patch $0\leq z<u<1$, the analytically continued expressions are valid for $0\leq u<z\leq 1$ (we always have $z\leq 1$ by definition) and therefore allow for the derivation of the factorization constraint. Using expansions in $1-\frac{z}{u}$ in (\ref{ConfBlocks}), this two point function can be shown to possess a finite limit at $u=z$. This has been anticipated in \cite{HoRi}. Moreover, since we are using analytic continuations, it must also be continuous at $u=z$. This feature has been postulated as an axiom in \cite{HoRi} and is what we referred to as the Hosomichi--Ribault proposal in the introduction. Our two point function shows all their requirements except the anticipated weakening of the Cardy--Lewellen factorization constraint, that is, the two point function in the patch $0\leq u<z\leq 1$ is completely determined from its expression in $0\leq z<u<1$, by analytic continuation.

\section{\label{Conclusion}Conclusions and Outlook}
We have argued that the boundary $\rm H_3^+$ model possesses a priori a variety of brane types, regular and irregular, discrete and continuous, that should all be analyzed, case by case, and checked for consistency. This programme is still far from being completed (see table \ref{T2} on the next page).\\
We derived $1/2$- and $b^{-2}/2$-shift equations in a systematic fashion for two ($\rho_1$ and $\rho_2$) out of four possible gluing maps (see chapter \ref{GlueConds} for a definition of the gluing maps). To be able to write down the $b^{-2}/2$-shift equations, we constructed a boundary two point function involving degenerate field with ${\rm sl}(2,\mathbb{C})$ label $b^{-2}/2$ and assumed that it could be analytically continued -- a point of view that has been discussed extensively in the introduction. We find that the known irregular continuous $AdS_2$ branes are consistent. Concerning the discrete $AdS_2$ branes, we show that only the regular ones with gluing map $\rho_1$ are consistent. For regular $\rho_2$, we can proof a no-solution-theorem. The known irregular discrete $AdS_2$ branes are found to be inconsistent with our second factorization constraint.\\
Regular discrete branes were studied in \cite{GKS}, however without considering the crucial $\sigma$-dependence of one point amplitude and coefficients in the shift equations. Here, we consider the full $\sigma$-dependence (that results in more complicated shift equations). We can give both factorization constraints (\ref{regAdS2d-rho1-Shift1}) and (\ref{regAdS2d-rho1-Shift2}) and their solution (\ref{regAdS2d-rho1-Sol}) explicitly.\\
About the regular continuous branes we could only be speculative. The reason is that the occuring residua of bulk-boundary OPE coefficients are not known explicitely. We also argue that they cannot be assumed to be identical to the corresponding expressions in the irregular case. Leaving these coefficients unfixed and starting from the most natural ansatz for a solution instead, we encounter rather unnatural requirements (see section \ref{regAdS2c}) that lead us to conjecture that these branes are not consistent.\\
In view of the Hosomichi--Ribault proposal, which anticipates a weakening of the Cardy-Lewellen constraints, our approach produces too strong constraints. But does this mean that we have to reject our analyticity assumption and discard the new constraints? We do not think so, because what we have demonstrated is that the assumption of analyticity is technically feasible and shows no unusual or even unphysical features. In fact, it does indeed lead to positive and beautiful statements: Irregular $AdS_2^{(c)}$ and regular $AdS_2^{(d)}$ branes are consistent with the analytic continuations that we study. This brane spectrum fits in very nicely with (a slight extension of) Cardy's and Ishibashi's results \cite{Cardy:BCsFusionAndVerlinde, Ishibashi:BoundaryAndCrosscapStates, SR1}. Moreover, two crucial properties of the Hosomichi--Ribault proposal are automatically satisfied by our analytic approach: Finiteness and continuity at $u=z$. Only the third property, which is weakening of the Cardy--Lewllen constraints, is not forced upon us when working analytically. Thus, working entirely inside the $\rm H_3^+$ model without a mapping to another theory, there does not seem to exist a good a priori reason why Cardy-Lewellen should be weakened. This, we think, is also a very interesting result of our work, besides confirming the consistency of known irregular $AdS_2^{(c)}$ branes and introducing a new type of consistent $AdS_2$ brane: The regular discrete solution.\\
\begin{table}
\begin{tabular}{|c||c|c|c|c|c|} \hline
             &$u$-dependence   
                       &\multicolumn{2}{c|}{shift equation (continuous)}   
                                 &\multicolumn{2}{c|}{shift equation (discrete)}\\
             &         
                       &for $\Theta_{1/2}?$      &for $\Theta_{b^{-2}/2}?$ 
                                 &for $\Theta_{1/2}?$      &for $\Theta_{b^{-2}/2}?$\\ \hline\hline
$\rho_1$     &$|u-\bar{u}|^{2j}$ 
                       &\cite{LOP}/\checkmark    &$\circledast$
                                 &$\circledast$            &$\circledast$\\ \cline{2-6}
             &$(u-\bar{u})^{2j}$ 
                       &\cite{GKS}/$\circledast$ &$\circledast$
                                 &\cite{GKS}/$\circledast$ &\cite{GKS}/$\circledast$\\ \hline
$\rho_2$     &$|u+\bar{u}|^{2j}$ 
                       &\cite{PST}/\checkmark    &\cite{Adorf:Continuous}/$\circledast$                 
                                 &\cite{SR1}/\checkmark    &\cite{Adorf:Continuous}/$\circledast$\\ \cline{2-6}
             &$(u+\bar{u})^{2j}$ 
                       &$\circledast$            &$\circledast$
                                 &$\circledast$            &$\circledast$\\ \hline
$\rho_3$     &$|-1+u\bar{u}|^{2j}$ 
                       &---                      &---                   
                                 &---                      &---\\ \cline{2-6}  
             &$(-1+u\bar{u})^{2j}$       
                       &---                      &---
                                 &\cite{GKS}               &\cite{GKS}\\ \hline
$\rho_4$     &$(1+u\bar{u})^{2j}$       
                       &---                      &---
                                 &\cite{PST}               &---\\ \hline                      
\end{tabular}
\caption{\label{T2} Classes of D-brane solutions: New contributions made in this paper are marked with a $\circledast$. Confirmed results are ticked $\checkmark$. Remember that we have reconsidered the results of \cite{GKS} for reasons explained in section \ref{BraneTypes} and that \cite{Adorf:Continuous} explored a different approach; see discussion in the text.}
\end{table}
Let us also remark that it is possible to construct a two point function which is finite and continuous along $u=z$, but at the same time not an analytic continuation, so that it has a chance of showing a weakened Cardy--Lewellen constraint. Such a two point function has been constructed recently by us, see \cite{Adorf:Continuous}. The irregular $AdS_2^{(d)}$ and $AdS_2^{(c)}$ branes are consistent with the constraints derived from this two point function.\footnote{The regular ones had not been treated.} Thus, together with this work, we have demonstrated that the branes of the $\rm H_3^+$ model show different properties, depending on whether the model is treated analytically or in accord with the Hosomichi--Ribault proposal. Remarkably however, the brane spectrum stays the same for both continuation prescriptions. The question that remains is what approach is more natural or if there is any reason to reject one of the two possibilities. We cannot decide this question here. All we can state is that when working entirely in the $\rm H_3^+$ model (i.e. without making any reference to results obtained from mapping to Liouville theory) and taking the point of view of shift equations, we do not see any reason why the Cardy--Lewellen constraint should be weakened.\\
What we like to mention is, that for the cigar CFT, a $b^{-2}/2$-shift equation for the $\mathrm{D1}^{(d)}$ and $\mathrm{D2}^{(d)}$ branes has been proposed in \cite{SR1} by analogy to $N=2$ Liouville Theory ($b^{-2}/2$-shift equations for $N=2$ Liouville theory have been derived in \cite{Hosomichi1}). These equations were indeed found to hold. This again points towards the pattern indicated above: Making reference to Liouville Theory (what in this situation probably also means a weakened Cardy--Lewellen constraint in the cigar CFT), the irregular discrete branes become consistent.\\
Now, one remark about the regular branes: We have remarked in section \ref{Regular-Irregular} that they decouple from the physical spectrum of closed $\rm H_3^+$ strings. Yet, these branes could still turn out to be important in view of string theory on $AdS_3$ or the cigar CFT (string in euclidean black hole background), because the physical spectrum of these theories is richer (see \cite{MaldaOoguri1} and \cite{DVV}, respectively). Another point is that in order to obtain the brane spectrum expected according to Cardy's and Ishibashi's work \cite{Cardy:BCsFusionAndVerlinde, Ishibashi:BoundaryAndCrosscapStates, SR1}, one has to include the regular branes when using the analytic continuation prescription. Moreover, one should not forget that a potentially interesting open string theory lives on the brane's worldvolume. It should not be forgotten just because no closed string scatters off the brane.\\
Besides our important observations about different types of branes in the $\rm H_3^+$ model, we also like to view this article as a step towards the classification of branes in that non--rational non--compact CFT. Next, one should collect results following the patterns continuous/discrete and regular/irregular more systematically, using two independent shift equations.\\
Looking at the progress made towards an understanding of the $\rm H_3^+$ model and Liouville theory, one can hope that more general non--rational CFTs will also be studied in the not too far future. Non--compact WZNW models with an ${\rm SL}(n,\mathbb{C})$ symmetry together with the ${\rm sl}(n)$ conformal Toda field theories \cite{FateevLitvinov:CTFT1} are natural next candidates. Indeed, it is highly desirable to get a grip on a larger variety of non--rational CFT models, as these provide the framework for a treatment of non--compact string backgrounds.

\vskip 2mm \noindent 
{\bf Acknowledgements:} We would like to thank J\"org Teschner for some helpful comments in the very early stages of the project. We are also grateful to Sylvain Ribault for a very constructive criticism of an earlier version of this paper. H.A. acknowledges financial support by the DFG-Graduiertenkolleg No. 282. The work of M.F. was partially supported by the European Union network HPRN-CT-2002-00325 (EUCLID).

\newpage
\appendix
\section{Exact Two Point Function Involving \boldmath$\Theta_{b^{-2}/2}$}
\subsection{\label{KniZa}Solution of the Knizhnik-Zamolodchikov Equation}
\subsubsection{\label{irrBranes}Irregular Branes}
In (\ref{irrB2pt-2}) we have given the general form of the two point function $G^{(2)}_{j,\alpha}(u_i\arrowvert z_i)$ fixed by the Ward identities. We use this expression in the Knizhnik-Zamolodchikov equation for $z_2$ which reads
\spliteq{-\frac{1}{b^2}\partial_{z_2}G^{(2)}_{j,\alpha}(u_i\arrowvert z_i)&=\\
\sum_a {\cal D}_{b^{-2}/2}^a(u_2) \otimes &\ebrac{\frac{{\cal D}_j^a(u_1)}{z_2-z_1}+\frac{\rho^{a}_{\hphantom{a}b}\bar{{\cal D}}_j^b(\bar{u}_1)}{z_2-\bar{z}_1}+\frac{\rho^{a}_{\hphantom{a}b}\bar{{\cal D}}_{b^{-2}/2}^b(\bar{u}_2)}{z_2-\bar{z}_2}}G^{(2)}_{j,\alpha}(u_i\arrowvert z_i).}
Mapping $z_1\rightarrow 0$, $\bar{z}_2\rightarrow 1$ and $\bar{z}_1\rightarrow\infty$ (i.e. $z_2\rightarrow z$) brings this equation to standard form
\spliteq{\label{KniZaStd}-&b^{-2}z(z-1)\partial_{z}H^{(2)}_{j,\alpha}(u\arrowvert z)=u(u-1)(u-z)\partial^2_u H^{(2)}_{j,\alpha}+\\
&+\set{\ebrac{1-2b^{-2}}u^2+\ebrac{b^{-2}-2j-2}uz+\ebrac{2j+b^{-2}}u+z}\partial_u H^{(2)}_{j,\alpha}+\\
&+\set{b^{-4}u+\ebrac{b^{-2}j-b^{-4}/2}z-b^{-2}j}H^{(2)}_{j,\alpha}.}
It is solved by (see \cite{JT1} and \cite{Appell}) $H^{(2)}_{j,\alpha}=\sum_{\epsilon=+,-,\times} a^{j}_{\epsilon}(\alpha){\cal F}^{s}_{j,\epsilon}$ with
\spliteq{\label{ConfBlocksTilde}{\cal F}^{s}_{j,+}(u\arrowvert z)&=z^{-j}(1-z)^{-b^{-2}/2}F_1(\alpha,\beta,\beta';\gamma\arrowvert u;z),\\
{\cal\tilde{F}}^{s}_{j,-}(u\arrowvert z)&=z^{\beta-\gamma+1-j}(1-z)^{\gamma-\alpha-1-b^{-2}/2}(u-z)^{-\beta}\cdot\\
&\cdot F_1\left(1-\beta',\beta,\alpha+1-\gamma;2+\beta-\gamma\left\arrowvert\frac{z}{z-u};\frac{z}{z-1}\right)\right.,\\
{\cal\tilde{F}}^{s}_{j,\times}(u\arrowvert z)&=z^{-j}(1-z)^{-b^{-2}/2}\e^{\i\pi(\alpha+1-\gamma)}\frac{\Gamma(\alpha)\Gamma(\gamma-\beta)}{\Gamma(\alpha+1-\beta)\Gamma(\gamma-1)}\cdot\\
&\cdot\left\{u^{-\alpha}F_1\left(\alpha,\alpha+1-\gamma,\beta';\alpha+1-\beta\left\arrowvert\frac{1}{u};\frac{z}{u}\right)\right.-\right.\\
&\left. -\e^{-\i\pi\alpha}\frac{\Gamma(\alpha+1-\beta)\Gamma(1-\gamma)}{\Gamma(\alpha+1-\gamma)\Gamma(1-\beta)}F_1(\alpha,\beta,\beta';\gamma\arrowvert u;z)\right\}.}
For our purposes, we like to replace the ${\cal\tilde{F}}^{s}_{j,-}$ block by
\spliteq{\label{Mod-Block}{\cal F}^{s}_{j,-}(u\arrowvert z)&=z^{-j}(1-z)^{-b^{-2}/2}u^{-\beta}z^{1+\beta-\gamma}\cdot\\
&\cdot F_1\left(1+\beta+\beta'-\gamma,\beta,1+\alpha-\gamma;2+\beta-\gamma\left\arrowvert\frac{z}{u};z\right)\right..}
This coincides with the one given in (\ref{ConfBlocksTilde}) in the overlap of their domains of convergence \cite{Appell} and thus, (\ref{Mod-Block}) is a continuation of the former ${\cal\tilde{F}}^{s}_{j,-}$ block. Also, we continue the first summand of ${\cal\tilde{F}}^{s}_{j,\times}$ to $(\frac{1}{u},\frac{z}{u}):=(\eta,\xi)\approx (\infty,0)$. Then, one of the resulting two terms precisely cancels the second summand of ${\cal\tilde{F}}^{s}_{j,\times}$ and we are only left with
\spliteq{\label{ModxBlock}{\cal F}^{s}_{j,\times}(u\arrowvert z)&=z^{-j}(1-z)^{-b^{-2}/2}u^{1-\gamma}\cdot\\
&\cdot G_2\left(\beta',1+\alpha-\gamma;1+\beta-\gamma,\gamma-1\left\arrowvert -\frac{z}{u};u\right.\right).}
With these modest improvements, the boundary two point function is now defined in the region $z<u$ ($u,z<1)$. For convenience, let us once and for all assemble the conformal blocks we are using:
\spliteq{\label{ConfBlocks}{\cal F}^{s}_{j,+}(u\arrowvert z)&=z^{-j}(1-z)^{-b^{-2}/2}F_1(\alpha,\beta,\beta';\gamma\arrowvert u;z),\\
{\cal F}^{s}_{j,-}(u\arrowvert z)&=z^{-j}(1-z)^{-b^{-2}/2}u^{-\beta}z^{1+\beta-\gamma}\cdot\\
&\cdot F_1\left(1+\beta+\beta'-\gamma,\beta,1+\alpha-\gamma;2+\beta-\gamma\left\arrowvert\frac{z}{u};z\right)\right.,\\
{\cal F}^{s}_{j,\times}(u\arrowvert z)&=z^{-j}(1-z)^{-b^{-2}/2}u^{1-\gamma}\cdot\\
&\cdot G_2\left(\beta',1+\alpha-\gamma;1+\beta-\gamma,\gamma-1\left\arrowvert -\frac{z}{u};u\right.\right).}
The functions $F_1(\alpha,\beta,\beta',\gamma\arrowvert u;z)$ and $G_2(\beta,\beta';\alpha,\alpha'\arrowvert u;z)$ are generalized hypergeometric functions: $F_1$ is the first one of Appell's double hypergeometric functions and $G_2$ is one of the functions appearing on Horn's list. We introduce them briefly in \ref{AppellHorn}. See the books \cite{Appell, Exton} for more information. For the occuring parameters we find
\eq{\alpha=\beta=-b^{-2},\hskip .3cm \beta'=-2j-1-b^{-2},\hskip .3cm \gamma=-2j-b^{-2}.}

\subsubsection{\label{regBranes}Regular Branes}
The whole procedure is just like before for the irregular branes. The solution is only slightly modified. It is given by
\spliteq{\label{ConfBlocksReg}{\cal F}^{s}_{j,+}(u\arrowvert z)&=z^{-j}(1-z)^{-j}F_1(\alpha,\beta,\beta';\gamma\arrowvert u;z),\\
{\cal F}^{s}_{j,-}(u\arrowvert z)&=z^{-j}(1-z)^{-j}u^{-\beta}z^{1+\beta-\gamma}\cdot\\
&\cdot F_1\left(1+\beta+\beta'-\gamma,\beta,1+\alpha-\gamma;2+\beta-\gamma\left\arrowvert\frac{z}{u};z\right)\right.,\\
{\cal F}^{s}_{j,\times}(u\arrowvert z)&=z^{-j}(1-z)^{-j}u^{1-\gamma}\cdot\\
&\cdot G_2\left(\beta',1+\alpha-\gamma;1+\beta-\gamma,\gamma-1\left\arrowvert -\frac{z}{u};u\right.\right),}
this time with parameters
\eq{\alpha=-2j,\hskip .3cm \beta=-b^{-2},\hskip .3cm \beta'=-2j-1-b^{-2},\hskip .3cm \gamma=-2j-b^{-2}.}
Note that the common $(1-z)$-dependence is changed here to $(1-z)^{-j}$.

\subsection{\label{LinComb}Finding the Correct Linear Combination of Conformal Blocks}
\subsubsection{\label{LinComb-irr2-1}Irregular Branes - Gluing Maps \boldmath$\rho_2$ and \boldmath$\rho_1$}
In order to obtain the exact result for the boundary two point function (\ref{irrB2pt-2}), all that is left to do is determine the coefficients $a^{j}_{\epsilon}(\alpha)$, i.e. find the correct linear combination of conformal blocks (\ref{ConfBlocks}). To this end, we use the operator product expansion (OPE) on the L.H.S. of (\ref{irrB2pt-2}) to obtain
\spliteq{&G^{(2)}_{j,\alpha}(u_1,u_2\arrowvert z_1,z_2)\simeq\\
&\hphantom{.}\simeq\abs{z_2-z_1}^{-2j}\abs{z_1-\bar{z}_1}^{-2h(j_+)}\abs{u_1\pm\bar{u}_1}^{2j+b^{-2}}C_+(j)A_{\sigma}(j_+\arrowvert\alpha)+\\
&\hphantom{.}+\abs{z_2-z_1}^{2j+2}\abs{u_2-u_1}^{2b^{-2}}\abs{z_1-\bar{z}_1}^{-2h(j_-)}\abs{u_1\pm\bar{u}_1}^{2j-b^{-2}}\cdot\\
&\hphantom{.+\abs{z_2-z_1}^{2j+2}\abs{u_2-u_1}^{2b^{-2}}\abs{z_1-\bar{z}_1}^{-2h(j_-)}.}\cdot C_-(j)A_{\sigma}(j_-\arrowvert\alpha)+\\
&\hphantom{.}+\abs{z_2-z_1}^{-2j}\abs{u_2-u_1}^{2(2j+1+b^{-2})}\abs{z_1-\bar{z}_1}^{-2h(j_{\times})}\abs{u_1\pm\bar{u}_1}^{-2j-2-b^{-2}}\cdot\\
&\hphantom{.+\abs{z_2-z_1}^{-2j}\abs{u_2-u_1}^{2(2j+1+b^{-2})}\abs{z_1-\bar{z}_1}^{-2h(j_{\times})}.}\cdot C_{\times}(j)A_{\sigma}(j_{\times}\arrowvert\alpha).}
We have used here that 
\spliteq{h(j_+)&\equiv h(j_{\times})=h(j)+h\brac{\frac{b^{-2}}{2}}-j\\
h(j_-)&=h(j)+h\brac{\frac{b^{-2}}{2}}+j+1.}
On the R.H.S. we can also take the limit $\abs{z_2-z_1}\rightarrow 0$ ($\Rightarrow z\rightarrow 0+$) followed by $\abs{u_2-u_1}\rightarrow 0$ ($\Rightarrow u\rightarrow 0+$). The conformal blocks (\ref{ConfBlocks}) behave as follows:
\spliteq{{\cal F}^{s}_{j,+}(u\arrowvert z)&\simeq z^{-j},\\
{\cal F}^{s}_{j,-}(u\arrowvert z)&\simeq z^{j+1}u^{b^{-2}},\\
{\cal F}^{s}_{j,\times}(u\arrowvert z)&\simeq z^{-j}u^{2j+1+b^{-2}}.}
Together with the prefactor
\spliteq{\abs{z_1-\bar{z}_1}&^{2[h(b^{-2}/2)-h(j)]}\abs{z_1-\bar{z}_2}^{-4h(b^{-2}/2)}\abs{u_1\pm\bar{u}_1}^{2j-b^{-2}}\abs{u_1\pm\bar{u}_2}^{2b^{-2}}\simeq\\
&\simeq\abs{z_1-\bar{z}_1}^{-2[h(b^{-2}/2)+h(j)]}\abs{u_1\pm\bar{u}_1}^{2j+b^{-2}}}
from (\ref{irrB2pt-2}) or (\ref{irrB2pt-1}) respectively, and recalling that 
\eq{z=\frac{\abs{z_2-z_1}^2}{\abs{z_2-\bar{z}_1}^2} \hskip .5cm {\rm and} \hskip .5cm
u=\frac{\abs{u_2-u_1}^2}{\abs{u_2\pm\bar{u}_1}^2},}
we find precisely
\eq{a^{j}_{\epsilon}(\alpha)=C_{\epsilon}(j)A_{\sigma}(j_{\epsilon}\arrowvert\alpha).}

\subsubsection{\label{LinComb-reg2}Regular Branes - Gluing Map \boldmath$\rho_2$}
Using the OPE on the L.H.S. of (\ref{regB2pt-2}), we find
\spliteq{&G^{(2)}_{j,\alpha}(u_1,u_2\arrowvert z_1,z_2)\simeq\\
&\hphantom{.}\simeq\abs{z_2-z_1}^{-2j}\brac{z_1-\bar{z}_1}^{-2h(j_+)}\brac{u_1+\bar{u}_1}^{2j+b^{-2}}C_+(j)A_{\sigma}(j_+\arrowvert\alpha)+\\
&\hphantom{.}+\abs{z_2-z_1}^{2j+2}\abs{u_2-u_1}^{2b^{-2}}\brac{z_1-\bar{z}_1}^{-2h(j_-)}\brac{u_1+\bar{u}_1}^{2j-b^{-2}}\\
&\hphantom{.+\abs{z_2-z_1}^{2j+2}\abs{u_2-u_1}^{2b^{-2}}\brac{z_1-\bar{z}_1}^{-2h(j_-)}.}\cdot C_-(j)A_{\sigma}(j_-\arrowvert\alpha)+\\
&\hphantom{.}+\abs{z_2-z_1}^{-2j}\abs{u_2-u_1}^{2(2j+1+b^{-2})}\brac{z_1-\bar{z}_1}^{-2h(j_{\times})}\brac{u_1+\bar{u}_1}^{-2j-2-b^{-2}}\\
&\hphantom{.+\abs{z_2-z_1}^{-2j}\abs{u_2-u_1}^{2(2j+1+b^{-2})}\brac{z_1-\bar{z}_1}^{-2h(j_{\times})}.}\cdot C_{\times}(j)A_{\sigma}(j_{\times}\arrowvert\alpha).}
Taking $\abs{z_2-z_1}\rightarrow 0$ ($\Rightarrow z\rightarrow 0-$) followed by $\abs{u_2-u_1}\rightarrow 0$ ($\Rightarrow u\rightarrow 0-$) on the R.H.S., the conformal blocks (\ref{ConfBlocksReg}) show the behaviour
\spliteq{{\cal F}^{s}_{j,+}(u\arrowvert z)&\simeq z^{-j},\\
{\cal F}^{s}_{j,-}(u\arrowvert z)&\simeq z^{j+1}u^{b^{-2}},\\
{\cal F}^{s}_{j,\times}(u\arrowvert z)&\simeq z^{-j}u^{2j+1+b^{-2}}.}
Remember that they are accompanied by the prefactor
\spliteq{\brac{z_1-\bar{z}_1}&^{-2h(j)}\brac{z_2-\bar{z}_2}^{-2h(b^{-2}/2)}\brac{u_1+\bar{u}_1}^{2j}\brac{u_2+\bar{u}_2}^{b^{-2}}\simeq\\
&\simeq\brac{z_1-\bar{z}_1}^{-2[h(j)+h(b^{-2/2})]}\brac{u_1+\bar{u}_1}^{2j+b^{-2}}}
from (\ref{regB2pt-2}) and that 
\eq{z=\frac{\abs{z_1-z_2}^2}{\brac{z_1-\bar{z}_1}\brac{z_2-\bar{z}_2}} \hskip .5cm {\rm and} \hskip .5cm
u=-\frac{\abs{u_1-u_2}^2}{\brac{u_1+\bar{u}_1}\brac{u_2+\bar{u}_2}}.}
Now, we need to be careful about phase factors that arise from $z^{-j}$, $u^{b^{-2}}$, and so on. In order to be consistent with the choice of phase we have to make because of the branch cut of the hypergeometric functions (that is, we take phases to be in $(-2\pi,0]$ - see section \ref{CloseLook}), we have to use the relations ($\nu\in\mathbb{C}$)
\spliteq{z^{\nu}&=\e^{-4\pi\i\nu}\abs{z_1-z_2}^{2\nu}\brac{z_1-\bar{z}_1}^{-\nu}\brac{z_2-\bar{z}_2}^{-\nu},\\
u^{\nu}&=\e^{\i\pi(\sigma-2)\nu}\abs{u_1-u_2}^{2\nu}\brac{u_1+\bar{u}_1}^{-\nu}\brac{u_2+\bar{u}_2}^{-\nu},}
where $\sigma=\sgn(u_1+\bar{u}_1)=\sgn(u_2+\bar{u}_2)$. One can check that this is correct by comparing the complex phases on both sides of the equations. With the help of this, it is quite obvious to see that the coefficients $a^{j,b^{-2}/2}_{\epsilon}$ need to be defined with phases just as in (\ref{a-coeffs-reg-2}).

\subsubsection{\label{LinComb-reg1}Regular Branes - Gluing Map \boldmath$\rho_1$}
Just as before, using the OPE on the L.H.S. of (\ref{regB2pt-1}), we find
\spliteq{&G^{(2)}_{j,\alpha}(u_1,u_2\arrowvert z_1,z_2)\simeq\\
&\hphantom{.}\simeq\abs{z_2-z_1}^{-2j}\brac{z_1-\bar{z}_1}^{-2h(j_+)}\brac{u_1-\bar{u}_1}^{2j+b^{-2}}C_+(j)A_{\sigma}(j_+\arrowvert\alpha)+\\
&\hphantom{.}+\abs{z_2-z_1}^{2j+2}\abs{u_2-u_1}^{2b^{-2}}\brac{z_1-\bar{z}_1}^{-2h(j_-)}\brac{u_1-\bar{u}_1}^{2j-b^{-2}}\\
&\hphantom{.+\abs{z_2-z_1}^{2j+2}\abs{u_2-u_1}^{2b^{-2}}\brac{z_1-\bar{z}_1}^{-2h(j_-)}.}\cdot C_-(j)A_{\sigma}(j_-\arrowvert\alpha)+\\
&\hphantom{.}+\abs{z_2-z_1}^{-2j}\abs{u_2-u_1}^{2(2j+1+b^{-2})}\brac{z_1-\bar{z}_1}^{-2h(j_{\times})}\brac{u_1-\bar{u}_1}^{-2j-2-b^{-2}}\\
&\hphantom{.+\abs{z_2-z_1}^{-2j}\abs{u_2-u_1}^{2(2j+1+b^{-2})}\brac{z_1-\bar{z}_1}^{-2h(j_{\times})}.}\cdot C_{\times}(j)A_{\sigma}(j_{\times}\arrowvert\alpha).}
Taking $\abs{z_2-z_1}\rightarrow 0$ ($\Rightarrow z\rightarrow 0-$) followed by $\abs{u_2-u_1}\rightarrow 0$ ($\Rightarrow u\rightarrow 0-$) on the R.H.S., the conformal blocks (\ref{ConfBlocksReg}) again show the behaviour
\spliteq{{\cal F}^{s}_{j,+}(u\arrowvert z)&\simeq z^{-j},\\
{\cal F}^{s}_{j,-}(u\arrowvert z)&\simeq z^{j+1}u^{b^{-2}},\\
{\cal F}^{s}_{j,\times}(u\arrowvert z)&\simeq z^{-j}u^{2j+1+b^{-2}}.}
This time they are accompanied by the prefactor
\spliteq{\brac{z_1-\bar{z}_1}&^{-2h(j)}\brac{z_2-\bar{z}_2}^{-2h(b^{-2}/2)}\brac{u_1-\bar{u}_1}^{2j}\brac{u_2-\bar{u}_2}^{b^{-2}}\simeq\\
&\simeq\brac{z_1-\bar{z}_1}^{-2[h(j)+h(b^{-2}/2)]}\brac{u_1-\bar{u}_1}^{2j+b^{-2}}}
from (\ref{regB2pt-1}) and
\eq{z=\frac{\abs{z_1-z_2}^2}{\brac{z_1-\bar{z}_1}\brac{z_2-\bar{z}_2}} \hskip .5cm {\rm and} \hskip .5cm
u=\frac{\abs{u_1-u_2}^2}{\brac{u_1-\bar{u}_1}\brac{u_2-\bar{u}_2}}.}
Again, we need to be careful about phase factors. We have to use ($\nu\in\mathbb{C}$)
\spliteq{z^{\nu}&=\e^{-4\pi\i\nu}\abs{z_1-z_2}^{2\nu}\brac{z_1-\bar{z}_1}^{-\nu}\brac{z_2-\bar{z}_2}^{-\nu},\\
u^{\nu}&=\e^{-\i\pi(\sigma+3)\nu}\abs{u_1-u_2}^{2\nu}\brac{u_1-\bar{u}_1}^{-\nu}\brac{u_2-\bar{u}_2}^{-\nu},}
where $\sigma=\sgn(u_1-\bar{u}_1)=\sgn(u_2-\bar{u}_2)$. With the help of this, it is easy to see that the coefficients $a^{j,b^{-2}/2}_{\epsilon}$ need to be defined with phases just as in (\ref{a-coeffs-reg-1}).

\section{Factorization Limit of the Exact Boundary Two Point Function}
\subsection{Irregular Branes}
We start from 
\spliteq{\label{ExB2pt}G^{(2)}_{j,\alpha}(u_1,u_2&\arrowvert z_1,z_2)=\abs{z_1-\bar{z}_1}^{2[h(b^{-2}/2)-h(j)]}\abs{z_1-\bar{z}_2}^{-4h(b^{-2}/2)}\cdot\\
&\cdot\abs{u_1\pm\bar{u}_1}^{2j-b^{-2}}\abs{u_1\pm\bar{u}_2}^{2b^{-2}}\sum_{\epsilon=+,-,\times}C_{\epsilon}(j)A_{\sigma}(j_{\epsilon}\arrowvert\alpha){\cal F}^{s}_{j,\epsilon}(u\arrowvert z),}
with conformal blocks ${\cal F}^{s}_{j,\epsilon}(u\arrowvert z)$ given in (\ref{ConfBlocks}). Since
\eq{1-z=\frac{4\Im(z_1)\Im(z_2)}{\abs{z_2-\bar{z}_1}^2},\hskip 1cm 1-u=\frac{(u_1\pm\bar{u}_1)(u_2\pm\bar{u}_2)}{\abs{u_2\pm\bar{u}_1}^2},}
taking the limit $\Im(z_2)\rightarrow 0$ followed by $(u_2\pm\bar{u}_2)\rightarrow 0$ implies $z\rightarrow 1-$ (the $z_i$ live in the upper half plane) and $u\rightarrow\rightarrow 1-$ (if $\sigma_1=\sigma_2$). Hence, the conformal blocks must be expanded in the variables $1-z$, $1-u$. We obtain these expansions of Appell's function $F_1$ and Horn's function $G_2$ by making repeated use of their one variable expansion, which have ordinary hypergeometric functions in the other variable as coefficients (see appendix \ref{AppellHorn}). To these occuring hypergeometric functions we employ standard analytic continuation formulae (collected in appendix \ref{Hypergeo}). Then we resum the series and, if necessary, repeat the procedure. We have given a detailed example in section \ref{CloseLook}. For now, let us just state the results here. For the irregular branes, note that the parameters $\alpha$, $\beta$, $\beta'$, $\gamma$ are not all independent of each other, but obey the relations $\alpha=\beta$ and $1+\beta'-\gamma=0$. We therefore eliminate $\alpha$ and $\beta'$ and only work with $\beta$ and $\gamma$. Up to terms which are of order $\set{1+{\cal O}(1-z,1-u)}$ we find that
\spliteq{{\cal F}^{s}_{j,+}(u\arrowvert z)&\simeq\frac{\Gamma(\gamma)\Gamma(\beta-1)}{\Gamma(\beta)\Gamma(\gamma-1)}(1-z)^{1+b^{-2}/2}(1-u)^{b^{-2}}+\\
&+\frac{\Gamma(\gamma)\Gamma(1-2\beta)}{\Gamma(\gamma-\beta)\Gamma(1-\beta)}(1-z)^{-b^{-2}/2}+\\
&+\frac{\Gamma(\gamma)\Gamma(1-\beta)\Gamma(2\beta-1)}{\Gamma(\gamma-\beta)\Gamma(\beta)\Gamma(\beta)}(1-z)^{-b^{-2}/2}(1-u)^{2b^{-2}+1},\\
{\cal F}^{s}_{j,-}(u\arrowvert z)&\simeq\frac{\Gamma(2+\beta-\gamma)\Gamma(\beta-1)}{\Gamma(\beta)\Gamma(1+\beta-\gamma)}\e^{\i\pi\beta}(1-z)^{1+b^{-2}/2}(1-u)^{b^{-2}}+\\
&+\frac{\Gamma(2+\beta-\gamma)\Gamma(1-2\beta)}{\Gamma(2-\gamma)\Gamma(1-\beta)\Gamma(1-\beta)\Gamma(\beta)}\e^{\i\pi\beta}(1-z)^{-b^{-2}/2}+\\
&+\frac{\Gamma(2+\beta-\gamma)\Gamma(2\beta-1)}{\Gamma(2-\gamma)\Gamma(\beta)\Gamma(\beta)\Gamma(\beta)}\e^{\i\pi\beta}(1-z)^{-b^{-2}/2}(1-u)^{2b^{-2}+1}+\\
&+\frac{\Gamma(2+\beta-\gamma)}{\Gamma(2-\gamma)\Gamma(\beta)}\e^{\i\pi\beta}(1-z)^{-b^{-2}/2}\sum_{n=0}^{\infty}h_n(\beta)\frac{(\beta)_n(\beta)_n}{(1)_n}\frac{u^n}{n!},}
\spliteq{{\cal F}^{s}_{j,\times}(u\arrowvert z)&\simeq\frac{\Gamma(2-\gamma)\Gamma(1-2\beta)\Gamma(\gamma-\beta)}{\Gamma(1-\beta)\Gamma(1-\beta)\Gamma(1-\beta)}(1-z)^{-b^{-2}/2}+\\
&+\left[\frac{\Gamma(2-\gamma)\Gamma(2\beta-\gamma)}{\Gamma(1+\beta-\gamma)\Gamma(1+\beta-\gamma)}\e^{\i\pi(\gamma-1)}-\right.\\
&\left. \hphantom{+[}-\frac{\Gamma(2-\gamma)\Gamma(\gamma-2\beta)\Gamma(2\beta-1)\Gamma(\gamma-\beta)}{\Gamma(\beta)\Gamma(1-\beta)\Gamma(1-\beta)\Gamma(\gamma-1)}\e^{\i\pi 2\beta}\right]\cdot\\
&\cdot (1-z)^{-b^{-2}/2}(1-u)^{2b^{-2}+1}.\nonumber}
Comparing to the bulk boundary OPE (\ref{BulkBdry2}), we see that the terms $\propto (1-z)^{1+b^{-2}/2}(1-u)^{b^{-2}}$ correspond to propagation of the identity. These are the terms that enter the shift equation. Note that the ${\cal F}^{s}_{j,\times}$ block does not contribute to these. The other terms which have a leading $z$-dependence $\propto (1-z)^{-b^{-2}/2}$ can be identified with propagation of the two other possible boundary fields $\Psi_{b^{-2}}$ (which has leading $u$-dependence $\propto (1-u)^{0}$) and $\Psi_{-b^{-2}-1}$ ($u$-dependence $\propto (1-u)^{2b^{-2}+1}$). (Recall that, because $\Theta_{b^{-2}/2}$ is a degenerate field, its bulk boundary OPE is highly restricted). Conveniently, all terms that appear fit in nicely with this interpretation. Only in the fourth summand in the block ${\cal F}^{s}_{j,-}$ we cannot extract the explicit $(1-u)$-dependence, because of the additional coefficients $h_n(\beta)$. They stem from the analytic continuation of a Gauss hypergeometric function in an exceptional (logarithmic) case - see section \ref{CloseLook} and apendix \ref{Hypergeo}. Yet, from its $(1-z)$-dependence it is clear that this term does not come from propagation of the identity.\\
Collecting the terms that stem from the identity propagation on both sides (use bulk boundary OPE on the L.H.S. and the above limit on the R.H.S. of (\ref{ExB2pt})), yields the desired shift equations (\ref{irrAdS2d-rho2-Shift2}) or (\ref{irrAdS2d-rho1-Shift2}) in the discrete and (\ref{irrAdS2c-rho2-1-Shift2}) in the continuous case.

\subsection{Regular Branes}
We start from 
\spliteq{\label{ExB2ptReg}G^{(2)}_{j,\alpha}(u_1,u_2&\arrowvert z_1,z_2)=\brac{z_1-\bar{z}_1}^{-2h(j)}\brac{z_2-\bar{z}_2}^{-2h(b^{-2}/2)}\cdot\\
&\cdot \brac{u_1\pm\bar{u}_1}^{2j}\brac{u_2\pm\bar{u}_2}^{b^{-2}}\sum_{\epsilon=+,-,\times}C_{\epsilon}(j)A_{\sigma}(j_{\epsilon}\arrowvert\alpha){\cal F}^{s}_{j,\epsilon}(u\arrowvert z),} 
with conformal blocks ${\cal F}^{s}_{j,\epsilon}(u\arrowvert z)$ given in (\ref{ConfBlocksReg}). Note that, for both gluing maps, the crossing ratios
\eq{z=\frac{\abs{z_1-z_2}^2}{\brac{z_1-\bar{z}_1}\brac{z_2-\bar{z}_2}} \hskip .5cm {\rm and} \hskip .5cm
u=\mp\frac{\abs{u_1-u_2}^2}{\brac{u_1\pm\bar{u}_1}\brac{u_2\pm\bar{u}_2}}}
take values in $z\in(-\infty,0]$ and $u\in(-\infty,0]$ (assuming $\sigma_1=\sigma_2$). In the limit $\Im(z_2)\rightarrow 0$, $(u_2\pm\bar{u}_2)\rightarrow 0$, we have this time that $z\rightarrow -\infty$ and $u\rightarrow\rightarrow -\infty$ ($\sigma_1=\sigma_2$). Hence, the conformal blocks must be expanded in the variables $\frac{1}{z}$, $\frac{1}{u}$. As before, we obtain these expansions of Appell's function $F_1$ and Horn's function $G_2$ by making repeated use of their one variable expansion, which have ordinary hypergeometric functions in the other variable as coefficients (see appendix \ref{AppellHorn}) to which we can apply standard analytic continuation formulae (collected in appendix \ref{Hypergeo}), resum and repeat everything if necessary - see section \ref{CloseLook} for an example of the procedure. Here we just state the results: For the regular branes, note that the parameters $\alpha$, $\beta$, $\beta'$, $\gamma$ are once again not all independent of each other, but obey the relations $\alpha+\beta-\gamma=0$ and $1+\beta'-\gamma=0$, which we use to eliminate $\alpha$ and $\beta'$ and only work with $\beta$ and $\gamma$. Up to terms which are of order $\set{1+{\cal O}(\frac{1}{z},\frac{1}{u})}$ we find that
\spliteq{{\cal F}^{s}_{j,+}(u\arrowvert z)&\simeq\left[\frac{\Gamma(\gamma)\Gamma(2\beta-\gamma)\Gamma(1-2\beta+\gamma)\Gamma(\beta-1)}{\Gamma(\beta)\Gamma(\beta)\Gamma(\gamma-1)\Gamma(1-\beta)}\right. +\\
&+\left.\frac{\Gamma(\gamma)\Gamma(\gamma-2\beta)\Gamma(1+2\beta-\gamma)\Gamma(\beta-1)}{\Gamma(\gamma-\beta)\Gamma(\gamma-1)\Gamma(\beta)\Gamma(1+\beta-\gamma)}\e^{\i\pi (2\beta-\gamma)}\right] \e^{-\i\pi j}z^{0}u^{0}+\\
&+\left[\frac{\Gamma(\gamma)\Gamma(1-2\beta)}{\Gamma(\gamma-\beta)\Gamma(1-\beta)}\e^{\i\pi (2j+1+2b^{-2})}\right]\e^{-\i\pi j}z^{1+b^{-2}}u^{b^{-2}}+\\
&+\left[\frac{\Gamma(\gamma)\Gamma(2\beta-\gamma)\Gamma(1-2\beta+\gamma)\Gamma(1-\beta)}{\Gamma(\beta)\Gamma(\beta)\Gamma(\gamma-\beta)\Gamma(2-2\beta)}\e^{\i\pi (\beta-1)}\right. +\\
&+\left.\frac{\Gamma(\gamma)\Gamma(\gamma-2\beta)\Gamma(2\beta-1)\Gamma(1+2\beta-\gamma)\Gamma(1-\beta)}{\Gamma(\gamma-\beta)\Gamma(\gamma-1)\Gamma(\beta)\Gamma(\beta)\Gamma(2-\gamma)}\e^{-\i\pi (1+\gamma-3\beta)}\right]\cdot\\
&\hphantom{+[}\cdot\e^{-\i\pi j}z^{1+b^{-2}}u^{-1-b^{-2}},\\
{\cal F}^{s}_{j,-}(u\arrowvert z)&\simeq\left[\frac{\Gamma(2+\beta-\gamma)\Gamma(\beta-1)}{\Gamma(\beta)\Gamma(1+\beta-\gamma)}\e^{-\i\pi (1-2\beta)}\right]\e^{-\i\pi j}z^{0}u^{0}+\\
&+\left[\frac{\Gamma(2+\beta-\gamma)\Gamma(1-2\beta)}{\Gamma(1-\beta)\Gamma(2-\gamma)}\e^{-\i\pi\beta}\right]\e^{-\i\pi j}z^{1+b^{-2}}u^{b^{-2}}+\\
&+\left[\frac{\Gamma(2+\beta-\gamma)\Gamma(2\beta-1)\Gamma(1-\beta)}{\Gamma(\beta)\Gamma(\beta)\Gamma(2-\gamma)}\e^{-\i\pi (2-3\beta)}\right]\e^{-\i\pi j}z^{1+b^{-2}}u^{-1-b^{-2}},\\
{\cal F}^{s}_{j,\times}(u\arrowvert z)&\simeq\left[\frac{\Gamma(2-\gamma)\Gamma(\gamma-2\beta)\Gamma(\gamma-\beta)\Gamma(1+2\beta-\gamma)\Gamma(\beta-1)}{\Gamma(1-\beta)\Gamma(1-\beta)\Gamma(\gamma-1)\Gamma(\beta)\Gamma(1+\beta-\gamma)}\e^{-\i\pi (1-2\beta)}\right. +\\
&+\left.\frac{\Gamma(2-\gamma)\Gamma(2\beta-\gamma)\Gamma(1-2\beta+\gamma)\Gamma(\beta-1)}{\Gamma(1+\beta-\gamma)\Gamma(1+\beta-\gamma)\Gamma(\gamma-1)\Gamma(1-\beta)}\e^{-\i\pi (1-\gamma)}\right] \e^{-\i\pi j}z^{0}u^{0}+\\
&+\left[\frac{\Gamma(2-\gamma)\Gamma(\gamma-\beta)\Gamma(1-2\beta)}{\Gamma(1-\beta)\Gamma(1-\beta)\Gamma(1-\beta)}\e^{-\i\pi\beta}\right]\e^{-\i\pi j}z^{1+b^{-2}}u^{b^{-2}}+\\
&+\left[\frac{\Gamma(\gamma-2\beta)\Gamma(\gamma-\beta)\Gamma(2\beta-1)\Gamma(1+2\beta-\gamma)\Gamma(1-\beta)}{\Gamma(1-\beta)\Gamma(1-\beta)\Gamma(\gamma-1)\Gamma(\beta)\Gamma(\beta)}\e^{-\i\pi (2-3\beta)}\right. +\\
&+\left.\frac{\Gamma(2-\gamma)\Gamma(2\beta-\gamma)\Gamma(1-2\beta+\gamma)\Gamma(1-\beta)}{\Gamma(1+\beta-\gamma)\Gamma(1+\beta-\gamma)\Gamma(\gamma-\beta)\Gamma(2-2\beta)}\e^{-\i\pi (2-\beta-\gamma)}\right]\cdot\\
&\hphantom{+[}\cdot\e^{-\i\pi j}z^{1+b^{-2}}u^{-1-b^{-2}}.}
We have written out these long and tedious terms for that the reader appreciate that all terms that arise at leading order are again grouped into three different asymptotics: $z^{0}u^{0}$ (corresponding to the propagating identity), $z^{1+b^{-2}}u^{b^{-2}}$ (corresponding to the field $\Psi_{b^{-2}}$) and $z^{1+b^{-2}}u^{-1-b^{-2}}$ (corresponding to $\Psi_{-b^{-2}-1}$). Presumably, all sums can be simplified. We have only done so for the identity contributions, because this is all we need. Writing down the identity contributions only and simplifying the occuring terms, the result looks much more convenient:
\spliteq{{\cal F}^{s}_{j,+}(u\arrowvert z)&\simeq\frac{2j+1+b^{-2}}{1+b^{-2}}\e^{\i\pi j}+\dots\,,\\
{\cal F}^{s}_{j,-}(u\arrowvert z)&\simeq -\frac{2j+1}{1+b^{-2}}\e^{-\i\pi (j+1-2b^{-2})}+\dots\,,\\
{\cal F}^{s}_{j,\times}(u\arrowvert z)&\simeq\frac{\Gamma(2j+2+b^{-2})\Gamma(-1-b^{-2})\Gamma(-2j)}{\Gamma(-2j-1-b^{-2})\Gamma(1+b^{-2})\Gamma(2j+1)}\e^{-\i\pi (3j+2b^{-2})}+\dots\,.}
The dots now represent the contributions of the two other fields that are different from the identity. Using the bulk boundary OPE for $\Theta_{b^{-2}/2}$ (\ref{BulkBdry2}) on the L.H.S. of (\ref{ExB2ptReg}), the leading contribution of the identity is
\spliteq{G^{(2)}_{j,\alpha}(u_1,u_2\arrowvert z_1,z_2)&\simeq\brac{z_2-\bar{z}_2}^{1+b^{-2}/2}\brac{u_2\pm\bar{u}_2}^{b^{-2}}\brac{z_1-\bar{z}_1}^{-2h(j)}\cdot\\
&\cdot\brac{u_1\pm\bar{u}_1}^{2j} C_{\sigma}(b^{-2}/2,0|\alpha)A_{\sigma}(j|\alpha)}
(or with $C_{\sigma}(b^{-2}/2,0|\alpha)$ replaced by $\tilde{c}_{\sigma}(b^{-2}/2,0|\alpha)$ for continuous branes). Now, the R.H.S. of (\ref{ExB2ptReg}) projected to the leading identity contribution is
\spliteq{G^{(2)}_{j,\alpha}(u_1,u_2\arrowvert z_1,z_2)&\simeq\brac{z_2-\bar{z}_2}^{1+b^{-2}/2}\brac{u_2\pm\bar{u}_2}^{b^{-2}}\brac{z_1-\bar{z}_1}^{-2h(j)}\cdot\\
&\cdot\brac{u_1\pm\bar{u}_1}^{2j}{\cal P}_{\mathbbm{1}}\lim_{z,u\rightarrow -\infty}\set{{\cal F}^{s}_{j,+}+{\cal F}^{s}_{j,-}+{\cal F}^{s}_{j,\times}}.}
Comparing the two expressions, we arrive at the shift equations (\ref{regAdS2d-rho2-Shift2}) and (\ref{regAdS2d-rho1-Shift2}) or, in the continuous case, (\ref{regAdS2c-rho2-Shift2}) and (\ref{regAdS2c-rho1-Shift2}).

\section{Bulk-Boundary OPE}
In this appendix we give the explicit form of the specific bulk-boundary OPEs needed in the calculations. For convenience, let us write the cases of gluing maps $\rho_1$ and $\rho_2$ in one formula. Also, we just write down the case of discrete open string spectrum, as the continuous case is easily obtained by changing $C_{\sigma}$ to $\tilde{c}_{\sigma}$. See also section \ref{Discrete-Continuous}, where we have introduced the generic bulk-boundary OPE and discussed the difference between discrete and continuous branes. Also note that further difference has to be made between the cases of regular and irregular branes. The formulae given below work for the irregular case, whereas for the discrete case, we need to replace the modulus $\abs{\dots}$ by ordinary brackets $\brac{\dots}$. This is necessary to ensure that the identification $C_{\sigma}=A_{\sigma}$ still holds true.

\subsection{\label{BulkBdryOPE1}Bulk-Boundary OPE for \boldmath$\Theta_{1/2}$}
\spliteq{\Theta_{1/2}(u_2|z_2)&=\abs{z_2-\bar{z}_2}^{\frac{3}{2}b^{2}}\abs{u_2\pm\bar{u}_2}C_{\sigma}(1/2,0|\alpha)\mathbbm{1}\set{1+{\cal O}\brac{z_2-\bar{z}_2}}+\\
&+\abs{z_2-\bar{z}_2}^{-\frac{1}{2}b^{2}}\abs{u_2\pm\bar{u}_2}^{2}C_{\sigma}(1/2,1|\alpha)\times\\
&\hphantom{+\abs{z_2-\bar{z}_2}^{-\frac{1}{2}b^{2}}}\times
\brac{{\cal J}\Psi}^{\alpha\,\alpha}_{1}\brac{u_2\left|{\rm Re}(z)\right.}\set{1+{\cal O}\brac{z_2-\bar{z}_2}}\,.}
The upper sign corresponds to gluing map $\rho_2$, the lower sign to $\rho_1$.

\subsection{\label{BulkBdryOPE2}Bulk-Boundary OPE for \boldmath$\Theta_{b^{-2}/2}$}
\spliteq{\label{BulkBdry2}\Theta_{b^{-2}/2}(u_2|z_2)&=\abs{z_2-\bar{z}_2}^{1+b^{-2}/2}\abs{u_2\pm\bar{u}_2}^{b^{-2}}C_{\sigma}(b^{-2}/2,0|\alpha)\mathbbm{1}\set{1+{\cal O}\brac{z_2-\bar{z}_2}}+\\
&+\abs{z_2-\bar{z}_2}^{-b^{-2}/2}\abs{u_2\pm\bar{u}_2}^{2b^{-2}+1}C_{\sigma}(b^{-2}/2,b^{-2}|\alpha)\times\\
&\hphantom{+\abs{z_2-\bar{z}_2}^{-b^{-2}/2}}\times
\brac{{\cal J}\Psi}^{\alpha\,\alpha}_{b^{-2}}\brac{u_2\left|{\rm Re}(z)\right.}\set{1+{\cal O}\brac{z_2-\bar{z}_2}}+\\
&+\abs{z_2-\bar{z}_2}^{-b^{-2}/2}C_{\sigma}(b^{-2}/2,-b^{-2}-1|\alpha)\times\\
&\hphantom{+\abs{z_2-\bar{z}_2}^{-b^{-2}/2}}\times
\brac{{\cal J}\Psi}^{\alpha\,\alpha}_{-b^{-2}-1}\brac{u_2\left|{\rm Re}(z)\right.}\set{1+{\cal O}\brac{z_2-\bar{z}_2}}\,.}
Again, upper sign corresponds to gluing map $\rho_2$, lower sign to $\rho_1$.

\section{Bulk OPE Coefficients}
We obtain the bulk OPE coefficients from the structure constants that were given in \cite{JT2}. We only need to be careful about the different normalisations of field operators. In \cite{JT2}, the operators $\phi_j(u|z)$ are used, whereas here (as well as in \cite{PST}) we are working with $\Theta_j(u|z):=B^{-1}(j)\phi_j(u|z)$, where $B(j)=(2j+1)R(j)/\pi$, $R(j)$ being the reflection amplitude, see (\ref{Rj}). With this, the structure constants $D(j,j_1,j_2)$ of \cite{JT2} have to be "dressed" by some factors of $B^{-1}$:
\eq{C(j,j_1,j_2):=D(j,j_1,j_2)B^{-1}(j_1)B^{-1}(j_2).}

\subsection{\label{OPE1}Bulk OPE Coefficients for the OPE with $\Theta_{1/2}$}
For completeness we give the bulk OPE coefficients with the degenerate field $\Theta_{1/2}$, although they are also written in \cite{PST}, using the same normalisation as we do. Since $\Theta_{1/2}$ is degenerate, the OPE is highly restricted. Only the field operators with $j_+=j+1/2$ and $j_-=j-1/2$ do occur. The corresponding coefficients are
\eq{C_+(j)=1, \hskip .5cm C_-(j)=\frac{1}{\nu_b}\frac{\Gamma(-b^2(2j+1))\Gamma(1+2b^2j)}{\Gamma(1+b^2(2j+1))\Gamma(-2b^2j)}.}

\subsection{\label{OPE2}Bulk OPE Coefficients for the OPE with $\Theta_{b^{-2}/2}$}
The singular vector labelled by $b^{-2}/2$ restricts the possibly occuring field operators in the operator product to those with labels $j_+:=j+b^{-2}/2$, $j_-:=j-b^{-2}/2$, $j_{\times}:=-j-1-b^{-2}/2$. The corresponding OPE coefficients can be easily calculated. We obtain
\spliteq{C_+(j)=1&, \hskip .5cm C_-(j)=-\nu_b^{-b^{-2}}\ebrac{b^2(2j+1)}^{-2},\\ 
C_{\times}(j)=-\frac{\nu_b^{-2j-1-b^{-2}}}{b^4}&\frac{\Gamma(1+b^{-2})}{\Gamma(1-b^{-2})}\frac{\Gamma(1+2j)\Gamma(-1-2j-b^{-2})\Gamma(-b^2(2j+1))}{\Gamma(-2j)\Gamma(2+2j+b^{-2})\Gamma(1+b^2(2j+1))}.}

\section{\label{RefCnstr}A Further Constraint on the One Point Amplitude from Reflection Symmetry}
\subsection{The Irregular One Point Amplitudes}
Due to the reflection symmetry (\ref{RefSymm}), the one point amplitude has to obey
\spliteq{\frac{\pi}{2j+1}&\abs{u\mp\bar{u}}^{2j}A_{\sigma}(j\arrowvert\alpha)=\\
&=-R(-j-1)\int_{\mathbb{C}}\d^2u'\abs{u-u'}^{4j}\abs{u'\mp\bar{u}'}^{-2j-2}A_{\sigma'}(-j-1\arrowvert\alpha).}
The upper sign corresponds to gluing map $\rho_1$, the lower sign to $\rho_2$. Note that $\sigma '\equiv\sigma(u')$. Since we can always expand $A_{\sigma '}(-j-1\arrowvert\alpha)=A^0(-j-1\arrowvert\alpha)+\sigma 'A^1(-j-1\arrowvert\alpha)$, we need to compute the integrals ($\epsilon\in\set{0,1}$):
\eq{I^{\mp}_{\epsilon}:=\int_{\mathbb{C}}\d^2u'\abs{u-u'}^{4j}\abs{u'\mp\bar{u}'}^{-2j-2}(\sigma ')^{\epsilon}.}

\subsubsection{Gluing Map \boldmath$\rho_1$ - Calculation of \boldmath$I^{-}_{\epsilon}$}
Assume $u_2>0$. We split the integral into
\spliteq{I^{-}_{\epsilon}&=(-)^{\epsilon}\int_{-\infty}^{+\infty}\d u'_1\int_{-\infty}^{0}\d u'_2\ebrac{(u_1-u'_1)^2+(u_2-u'_2)^2}^{2j}(-2u'_2)^{-2j-2}+\\
&+\int_{-\infty}^{+\infty}\d u'_1\int_{0}^{u_2}\d u'_2\ebrac{(u_1-u'_1)^2+(u_2-u'_2)^2}^{2j}(2u'_2)^{-2j-2}+\\
&+\int_{-\infty}^{+\infty}\d u'_1\int_{u_2}^{+\infty}\d u'_2\ebrac{(u_1-u'_1)^2+(u_2-u'_2)^2}^{2j}(2u'_2)^{-2j-2}\\
&\equiv (-)^{\epsilon}I_1^{>}+I_2^{>}+I_3^{>}.}
Being careful about signs and using some Gamma function identities (see \ref{Gamma}), we obtain
\eq{I_1^{>}=-\frac{\pi}{2j+1}\abs{u-\bar{u}}^{2j},\hskip .5cm I_2^{>}=-I_3^{>}.}
Now, assume $u_2<0$. In this case, we choose the following splitting
\spliteq{I^{-}_{\epsilon}&=(-)^{\epsilon}\int_{-\infty}^{+\infty}\d u'_1\int_{-\infty}^{u_2}\d u'_2\ebrac{(u_1-u'_1)^2+(u_2-u'_2)^2}^{2j}(-2u'_2)^{-2j-2}+\\
&+(-)^{\epsilon}\int_{-\infty}^{+\infty}\d u'_1\int_{u_2}^{0}\d u'_2\ebrac{(u_1-u'_1)^2+(u_2-u'_2)^2}^{2j}(-2u'_2)^{-2j-2}+\\
&+\int_{-\infty}^{+\infty}\d u'_1\int_{0}^{+\infty}\d u'_2\ebrac{(u_1-u'_1)^2+(u_2-u'_2)^2}^{2j}(2u'_2)^{-2j-2}\\
&\equiv (-)^{\epsilon}I_1^{<}+(-)^{\epsilon}I_2^{<}+I_3^{<}.}
This time we get
\eq{I_1^{<}=-I_2^{<},\hskip .5cm I_3^{<}=-\frac{\pi}{2j+1}\abs{u-\bar{u}}^{2j}.}
Assembling, we obtain
\eq{I^{-}_{\epsilon}=-\frac{\pi}{2j+1}\abs{u-\bar{u}}^{2j}(-\sigma)^{\epsilon}.}

\subsubsection{Gluing Map \boldmath$\rho_2$ - Calculation of \boldmath$I^{+}_{\epsilon}$}
Splitting the integral as before and renaming the integration variables, it is easy to see that
\eq{I^{+}_{\epsilon}=I^{-}_{\epsilon}(u_1\leftrightarrow u_2)=-\frac{\pi}{2j+1}\abs{u+\bar{u}}^{2j}(-\sigma)^{\epsilon}.}

\subsubsection{The Constraint for the Irregular One Point Amplitudes}
Putting things together, we arrive at the constraint
\eq{A_{\sigma}(j\arrowvert\alpha)=R(-j-1)A_{-\sigma}(-j-1\arrowvert\alpha).}
Using the definition of the reflection amplitude (\ref{Rj}), we are led to redefine the one point amplitude
\eq{f_{\sigma}(j):=\nu _b^j\Gamma(1+b^2(2j+1))A_{\sigma}(j\arrowvert\alpha)}
(note that we have dropped the $\alpha$-dependence of $f_{\sigma}$). For this redefined one point amplitude, the constraint simply reads
\eq{\label{RefSymmCnstr-irr}f_{\sigma}(j)=-f_{-\sigma}(-j-1).}

\subsection{The Regular One Point Amplitudes}
This time we need to compute the integrals ($\epsilon\in\set{0,1}$):
\eq{I^{\mp}_{\epsilon}:=\int_{\mathbb{C}}\d^2u'\abs{u-u'}^{4j}\brac{u'\mp\bar{u}'}^{-2j-2}(\sigma ')^{\epsilon}.} Up to a sign, the result is very much the same as before:
\eq{I^{\mp}_{\epsilon}=\frac{\pi}{2j+1}\brac{u\mp\bar{u}}^{2j}(-\sigma)^{\epsilon}.}
Therefore, in the regular case, the constraint for the redefined one point amplitude is
\eq{\label{RefSymmCnstr-reg}f_{\sigma}(j)=+f_{-\sigma}(-j-1).}

\section{\label{NoSolution}A No-Solution-Theorem}
In this appendix we give details of the proof that there is no solution to both factorization constraints together with the reflection symmetry constraint in the case of regular discrete branes with gluing map $\rho_2$ (see section \ref{regAdS2d-rho2}). Let us make the redefinition 
\eq{f_{\sigma}(j)=:-\frac{\pi\e^{\i\frac{\pi}{4}b^2}}{\Gamma(-b^2)}\frac{\e^{-\i\pi\frac{b^2}{4}(2j+1)^2}}{\sin[\pi b^2 (2j+1)]}g_{\sigma}(j)\nonumber}
and work with $g_{\sigma}(j)$ here. Note that it has opposite parity from $f_{\sigma}(j)$. The shift equations (\ref{regAdS2d-rho2-Shift1}) and (\ref{regAdS2d-rho2-Shift2}) in terms of $g_{\sigma}(j)$ are given as ${\bf (2)}$ and ${\bf (3)}$ in the following\vskip .3cm

\noindent {\sl Theorem:} The system of equations
\spliteq{&{\bf (1)}\hskip .2cm g_{\sigma}(j)=-g_{-\sigma}(-j-1)\\
&{\bf (2)}\hskip .2cm g_{\sigma}(1/2)g_{\sigma}(j)=g_{\sigma}(j+1/2)-g_{\sigma}(j-1/2)\\
&{\bf (3)}\hskip .2cm g_{\sigma}(b^{-2}/2)g_{\sigma}(j)=\e^{-\i\pi 4j}g_{\sigma}(j+b^{-2}/2)+\e^{\i\pi 4j}\e^{-\i\pi\sigma b^{-2}}g_{\sigma}(j-b^{-2}/2)-\\
&\hphantom{{\bf (3)}\hskip .2cm g_{\sigma}(b^{-2}/2)g_{\sigma}(j)=}-\e^{-\i\pi 4j}\e^{-\i\pi\sigma (2j+b^{-2})}g_{\sigma}(-j-1-b^{-2}/2)\nonumber}
does not admit an interpolating solution (in the sense of section \ref{regAdS2d-rho2}).\vskip .3cm

\noindent In order to proof this result, we proceed in two steps. The first one is to show that any functions satisfying ${\bf (1)}$ and ${\bf (2)}$ must be $1$-periodic or $1$-antiperiodic. The second step establishes that any $1$-periodic or $1$-antiperiodic function cannot satisfy ${\bf (3)}$.\vskip.3cm

\noindent {\sl Proof - 1st Step. Any solution to ${\bf (1)}$ and ${\bf (2)}$ must be periodic (antiperiodic) with period (antiperiod) $1$:} Take ${\bf (2)}$ at $j=-1/2$ and use $g_{\sigma}(-1)=-g_{-\sigma}(0)$ to obtain
\eq{g_{\sigma}(1/2)g_{\sigma}(-1/2)=g_{\sigma}(0)+g_{-\sigma}(0).\nonumber}
Doing the same for $\sigma\mapsto -\sigma$ and using $g_{-\sigma}(-1/2)=-g_{\sigma}(-1/2)$ yields
\eq{-g_{-\sigma}(1/2)g_{\sigma}(-1/2)=g_{-\sigma}(0)+g_{\sigma}(0).\nonumber}
Together, these equations imply
\eq{g_{\sigma}(-1/2)=0 \hskip .5cm \mathrm{or} \hskip .5cm g_{\sigma}(1/2)=-g_{-\sigma}(1/2).\nonumber}
{\sl ${\bf I)}$ Assume $g_{\sigma}(-1/2)=0$:} Taking ${\bf (2)}$ at $j=0$ then implies that either $g_{\sigma}(0)=1$ or $g_{\sigma}(1/2)=0$. But $g_{\sigma}(0)=1$ cannot be true, since in that case ${\bf (2)}$ for $j=-1/2$ tells us that $g_{\sigma}(-1)=g_{\sigma}(0)=1$. But at the same time, by ${\bf (1)}$, we must have $g_{\sigma}(0)=-g_{-\sigma}(-1)=-1$ which is a contradiction. Hence, $g_{\sigma}(1/2)=0$. But then, ${\bf (2)}$ reduces to the statement that $g_{\sigma}(j)$ is $1$-periodic.\newline
\noindent {\sl ${\bf II)}$ Assume $g_{\sigma}(1/2)=-g_{-\sigma}(1/2)$:} ${\bf (2)}$ at $j=0$ and using $g_{\sigma}(-1/2)=-g_{-\sigma}(-1/2)$ as well as the assumption gives
\eq{-g_{-\sigma}(1/2)g_{\sigma}(0)=-g_{-\sigma}(1/2)+g_{-\sigma}(-1/2).\nonumber}
Also, ${\bf (2)}$ with $j=0$ and $\sigma\mapsto -\sigma$ produces
\eq{g_{-\sigma}(1/2)g_{-\sigma}(0)=g_{-\sigma}(1/2)-g_{-\sigma}(-1/2).\nonumber}
Both equations together imply
\eq{g_{-\sigma}(1/2)=0 \hskip .5cm \mathrm{or} \hskip .5cm g_{\sigma}(0)=g_{-\sigma}(0).\nonumber}
{\sl ${\bf A)}$ Assume $g_{-\sigma}(1/2)=0$:} In this case, equation ${\bf (2)}$ again reduces to the $1$-periodicity of $g_{\sigma}(j)$.\newline
\noindent {\sl ${\bf B)}$ Assume $g_{\sigma}(0)=g_{-\sigma}(0)$:} An induction argument shows that
\eq{(*) \hskip .5cm g_{\sigma}(k/2)=(-)^k g_{-\sigma}(k/2), \hskip .5cm k\in\mathbb{Z}.\nonumber}
((This is clear for $k=-1,0,1$. Then the step is taken from knowledge of $\set{k-1,k}\mapsto k+1$ using ${\bf (2)}$ twice, once for $\sigma$ and once for $-\sigma$. The step from $\set{k+1,k}\mapsto k-1$ is done in the same way.)) Now, redefine 
\eq{g_{\sigma}(j)=:\i\sigma\e^{\i\frac{\pi}{2}\sigma (2j+1)}h_{\sigma}(j) \hskip .2cm \mathrm{with} \hskip .2cm h_{\sigma}(k/2)=h_{-\sigma}(k/2) \hskip .2cm \mathrm{for} \hskip .2cm k\in\mathbb{Z}.\nonumber}
Note that the prefactor $\exp[\i\frac{\pi}{2}\sigma (2j+1)]$ is $1$-antiperiodic.
Such an $h_{\sigma}(j)$ satisfies $(*)$ automatically as well as
\spliteq{&{\bf (1')}\hskip .2cm h_{\sigma}(j)=h_{-\sigma}(-j-1)\\
&{\bf (2')}\hskip .2cm \i\sigma h_{\sigma}(1/2)h_{\sigma}(j)=h_{\sigma}(j+1/2)-h_{\sigma}(j-1/2).\nonumber}
Now, taking ${\bf (2')}$ for $j=1/2$ and using $h_{\sigma}(k/2)=h_{-\sigma}(k/2)$, we gain the relation
\eq{\i\sigma\ebrac{h_{-\sigma}(1/2)}^2=h_{-\sigma}(1)-h_{-\sigma}(0),\nonumber}
while at the same time, ${\bf (2')}$ at $j=1/2$ and $\sigma\mapsto -\sigma$ yields
\eq{-\i\sigma\ebrac{h_{-\sigma}(1/2)}^2=h_{-\sigma}(1)-h_{-\sigma}(0).\nonumber} 
From these two equations we see immediately that $h_{-\sigma}(1/2)=0$ what implies that $h_{\sigma}(j)$ must be $1$-periodic. Taking the $1$-antiperiodic prefactor $\exp[\i\frac{\pi}{2}\sigma (2j+1)]$ from the last redefinition into account, we have thus established, that in this case $g_{\sigma}(j)$ is $1$-antiperiodic. This finishes the first part of our proof.\vskip .3cm

\noindent {\sl Proof - 2nd Step. A $1$-periodic or $1$-antiperiodic function cannot satisfy ${\bf (3)}$:} The central part is to establish $g_{\sigma}(b^{-2}/2)=0$. This has to be done separately for $1$-periodic and $1$-antiperiodic functions.\newline
\noindent {\sl ${\bf I)}$ $g_{\sigma}(b^{-2}/2)=0$ for $1$-periodic $g_{\sigma}(j)$:} Using ${\bf (3)}$ at $j=0$ and the $1$-periodicity yields
\eq{g_{\sigma}(b^{-2}/2)g_{\sigma}(0)=g_{\sigma}(b^{-2}/2).}
Hence, $g_{\sigma}(0)=1$ or $g_{\sigma}(b^{-2}/2)=0$, but $g_{\sigma}(0)=1$ cannot be true because of the $1$-periodicity: 
\eq{1=g_{\sigma}(0)=-g_{-\sigma}(-1)=-g_{-\sigma}(0)=-1 \hskip .5cm \nonumber}  %\lightning does not work ????
which is a contradiction.\newline
\noindent {\sl ${\bf II)}$ $g_{\sigma}(b^{-2}/2)=0$ for $1$-antiperiodic $g_{\sigma}(j)$:} Here, we use ${\bf (3)}$ for $j=1/2$ and ${\bf (2)}$ for $j=b^{-2}/2$. This results in
\eq{g_{\sigma}(1/2+b^{-2}/2)=2\e^{-\i\pi\sigma b^{-2}}g_{\sigma}(1/2-b^{-2}/2).\nonumber}
But now, ${\bf (1)}$ and $1$-antiperiodicity also tell us that $g_{\sigma}(1/2-b^{-2}/2)=-g_{-\sigma}(1/2+b^{-2}/2)$, so that this equation becomes
\eq{g_{\sigma}(1/2+b^{-2}/2)=-2\e^{-\i\pi\sigma b^{-2}}g_{-\sigma}(1/2+b^{-2}/2).\nonumber}
Using this relation twices reveals that $g_{\sigma}(1/2+b^{-2}/2)=0$ and therefore, ${\bf (2)}$ for $j=b^{-2}/2$ implies $g_{\sigma}(b^{-2}/2)=0$ ((recall that necessarily $g_{\sigma}(1/2)\neq 0$, because of $1$-antiperiodicity)).\newline
\noindent It is now establishes that $g_{\sigma}(b^{-2}/2)=0$ in either case. In the remaining part we can treat $1$-periodicity and $1$-antiperiodicity simultaneously as the only difference from here on is a sign difference. In the following, upper signs will correspond to $1$-periodicity, lower signs to $1$-antiperiodicity. Because of $g_{\sigma}(b^{-2}/2)=0$, equation ${\bf (3)}$ is now
\spliteq{(*)^2 \hskip .5cm 0=\e^{-\i\pi 4j}&g_{\sigma}(j+b^{-2}/2)+\e^{\i\pi 4j}\e^{-\i\pi\sigma b^{-2}}g_{\sigma}(j-b^{-2}/2)-\\
&-\e^{-\i\pi 4j}\e^{-\i\pi\sigma (2j+b^{-2})}g_{\sigma}(-j-1-b^{-2}/2).\nonumber}
Taking it at $j\mapsto -j$, using $1$-(anti)periodicity and multiplying by $\exp[-\i\pi\sigma 2j]$ produces
\spliteq{0=\e^{\i\pi 4j}\e^{-\i\pi\sigma 2j}&g_{\sigma}(-j+b^{-2}/2)\mp\e^{\i\pi 4j}\e^{-\i\pi\sigma b^{-2}}g_{\sigma}(j-b^{-2}/2)\pm\\
&\pm\e^{-\i\pi 4j}\e^{-\i\pi\sigma (2j+b^{-2})}g_{\sigma}(-j-1-b^{-2}/2).\nonumber}
Adding (if $1$-periodic) or subtracting (if $1$-antiperiodic) these two equations, we obtain
\eq{\e^{-\i\pi 4j}g_{\sigma}(j+b^{-2}/2)=\e^{\i\pi 4j}\e^{-\i\pi\sigma 2j}g_{-\sigma}(j-b^{-2}/2).\nonumber}
Plugging this back into $(*)^2$, we can finally derive the relation
\eq{g_{-\sigma}(j-b^{-2}/2)=-2\e^{\i\pi\sigma (2j-b^{-2})}g_{\sigma}(j-b^{-2}/2)\nonumber}
which, when made use of twice implies the claimed $g_{\sigma}(j)=0$ for all $j$. This concludes the proof of our no-solution-theorem.

\section{Some Useful Formulae}
\subsection{\label{Gamma}\boldmath$\Gamma$ Function Identities}
\eq{\int_0^1\mathrm{d}t\hskip .1cm t^{a-1}(1-t)^{b-1}=\frac{\Gamma(a)\Gamma(b)}{\Gamma(a+b)}}
\eq{\int_0^{\infty}\mathrm{d}t\hskip .1cm (1+t^2)^{\alpha}=\frac{\sqrt{\pi}}{2}\frac{\Gamma(-\alpha-\frac{1}{2})}{\Gamma(-\alpha)}}
\eq{\Gamma(2j)=\frac{1}{\sqrt{\pi}}(2)^{2j-1}\Gamma(j)\Gamma(j+\frac{1}{2})}
\eq{\Gamma(z)\Gamma(1-z)=\frac{\pi}{\sin(\pi z)}}

\subsection{\label{Pochhammer}Pochhammer Symbol and Identities} 
The Pochhammer symbol is defined to be
\eq{(\alpha)_m:=\frac{\Gamma(\alpha+m)}{\Gamma(\alpha)}\,.} 
From this definition and the functional equation of Euler's gamma function, $\alpha\Gamma(\alpha)=\Gamma(\alpha+1)$, one easily gets the following identites: 
\spliteq{(\alpha)_{-m}&=\frac{(-)^m}{(1-\alpha)_m}\,,\\
(\alpha)_{m+n}&=\left\{\begin{array}{l} (\alpha+m)_n (\alpha)_m\\ (\alpha+n)_m (\alpha)_n \end{array}\right.,\\ 
(\alpha)_{m-n}&=\left\{\begin{array}{l} (\alpha+m)_{-n} (\alpha)_m\\ (\alpha-n)_m (\alpha)_{-n} \end{array}\right..}

\subsection{\label{Hypergeo}Analytic Continuations of the Hypergeometric Function}
The formulae stated here are taken from \cite{Bateman}. Note that for the hypergeometric function to exist, we always need $c\notin\mathbb{Z}_{\leq 0}$.

\paragraph{Generic Case (\boldmath$b-a\notin\mathbb{Z}$):}
\spliteq{F\left(a,b;c\left\arrowvert\frac{1}{u}\right)\right.&=\frac{\Gamma(c)\Gamma(b-a)}{\Gamma(b)\Gamma(c-a)}\brac{-\frac{1}{u}}^{-a}F(a,1-c+a;1-b+a\arrowvert u)+\\
&+\frac{\Gamma(c)\Gamma(a-b)}{\Gamma(a)\Gamma(c-b)}\brac{-\frac{1}{u}}^{-b}F(b,1-c+b;1-a+b\arrowvert u).}

\paragraph{Generic Case (\boldmath$c-a-b\notin\mathbb{Z}$):}
\spliteq{F\left(a,b;c\left\arrowvert z\right)\right.=\frac{\Gamma(c)\Gamma(c-a-b)}{\Gamma(c-a)\Gamma(c-b)}F(a,&b;a+b-c+1\arrowvert 1-z)+\\
+\frac{\Gamma(c)\Gamma(a+b-c)}{\Gamma(a)\Gamma(b)}(1-z)^{c-a-b}&F(c-a,c-b;c-a-b+1\arrowvert 1-z).}

\paragraph{Logarithmic Case (\boldmath$b-a=:m\in\mathbb{Z}_{\geq 0}$):}
\spliteq{F\left(a,b;c\left\arrowvert\frac{1}{u}\right)\right.&=\frac{\Gamma(c)}{\Gamma(b)\Gamma(c-a)}\brac{-\frac{1}{u}}^{-b}\sum_{n=0}^{\infty}\frac{(a)_{n+m}(1-c+a)_{n+m}}{n!(n+m)!}\cdot\\
&\hphantom{=\frac{\Gamma(c)}{\Gamma(b)\Gamma(c-a)}\brac{-\frac{1}{u}}^{-b}\cdot}\cdot\brac{\frac{1}{u}}^{-n}\ebrac{\log\brac{-\frac{1}{u}}+h_n}+\\
&+\frac{\Gamma(c)}{\Gamma(b)}\brac{-\frac{1}{u}}^{-a}\sum_{n}^{m-1}\frac{\Gamma(m-n)(a)_n}{\Gamma(c-a-n)n!}\brac{\frac{1}{u}}^{-n}.}
Note that if $b-a\in\mathbb{Z}$, it is no restriction to take $b-a=m\in\mathbb{Z}_{\geq 0}$, as this can always be achieved by exchanging the r\^{o}les of $a$ and $b$ if necessary. The occuring $h_n\equiv h_n(a,c,m)$ is defined as
\eq{h_n(a,c,m):=\psi(1+m+n)+\psi(1+n)-\psi(a+m+n)-\psi(c-a-m-n),}
with $\psi(z)$ being the logarithmic derivative of the gamma function:
\eq{\psi(z):=\frac{\Gamma'(z)}{\Gamma(z)}.}

\subsection{\label{AppellHorn}Appell's Function \boldmath$F_1$ and Horn's Function \boldmath$G_2$}
\paragraph{Definition as Convergent Series:} \noindent The definition of Appell's function $F_1$
is \eq{F_1(\alpha,\beta,\beta';\gamma\arrowvert
u;z):=\sum_{m,n=0}^{\infty}\frac{(\alpha)_{m+n}(\beta)_m(\beta')_n}{(\gamma)_{m+n}}\frac{u^m}{m!}
\frac{z^n}{n!}\,.} It is convergent for complex $u$ and $z$ in the domain $\abs{u}<1$,
$\abs{z}<1$. For the third parameter $\gamma$ we need $\gamma\neq 0,-1,-2,\dots$. Horn's
function $G_2$ is defined by \eq{G_2(\beta,\beta';\alpha,\alpha'\arrowvert
u;z):=\sum_{m,n=0}^{\infty}(\beta)_m(\beta')_n(\alpha)_{n-m}(\alpha')_{m-n}\frac{u^m}{m!}\frac{z^
n}{n!}\,.} This series also converges for complex $u$ and $z$ with $\abs{u}<1$, $\abs{z}<1$. Its
parameters $\alpha$ and $\alpha'$ must be such that $\alpha\neq 1,2,3,\dots$ and $\alpha'\neq
1,2,3,\dots$. Both special functions are solutions to a certain system of partial differential
equations (see e.g. \cite{Exton}). This can be used to extend their definitions to domains
reaching outside $\abs{u}<1$, $\abs{z}<1$.

\paragraph{Generalized Series Representations:} \noindent Employing the Pochhammer symbol
identites stated in \ref{Pochhammer}, one deduces that
\eq{F_1(\alpha,\beta,\beta';\gamma\arrowvert u;z)=\sum_{n=0}^{\infty}\frac{(\alpha)_n
(\beta')_n}{(\gamma)_n}F(\alpha+n,\beta;\gamma+n\arrowvert u)\frac{z^n}{n!}\,,} $F$ being the
standard hypergeometric function. Of course, there is an analogous statment about the expansion
in the variable $u$. It is simply obtained by exchanging $\beta$ and $\beta'$ on the RHS.

\noindent The corresponding expansion for $G_2$ is obtained in the same manner and reads
\eq{G_2(\beta,\beta';\alpha,\alpha'\arrowvert u;z)=\sum_{n=0}^{\infty}\frac{(\alpha)_n
(\beta')_n}{(1-\alpha')_n}F(\alpha'-n,\beta;1-\alpha-n\arrowvert -u)\frac{(-z)^n}{n!}\,.} The
analogous expansion in the variable $u$ is obtained by exchanging $\alpha$ and
$\alpha'$ as well as $\beta$ and $\beta'$ on the RHS.

\paragraph{Analytic Continuations:}\noindent Using the expansions given in the previous paragraph together with the continuation formulae for Gauss' hypergeometric function, one deduces the analytic continuations of Appell's function $F_1$. The generic cases can be found in \cite{Exton}.

\bibliographystyle{utphys}
\bibliography{draftbib}

\end{document}